\documentclass{article}

\PassOptionsToPackage{numbers, compress}{natbib}

\usepackage[final]{neurips_2021}

\usepackage[utf8]{inputenc} 
\usepackage[T1]{fontenc}    
\usepackage{hyperref}       
\usepackage{url}            
\usepackage{booktabs}       
\usepackage{amsfonts}       
\usepackage{nicefrac}       
\usepackage{microtype}      
\usepackage{xcolor}         
\usepackage{graphicx}
\usepackage{amsmath}
\usepackage{adjustbox}
\usepackage{caption}
\usepackage{subcaption}
\usepackage[toc,page]{appendix}
\title{Deep Contextual  Video Compression}

\author{%
    Jiahao Li, Bin Li, Yan Lu\\
    Microsoft Research Asia\\
    \{li.jiahao, libin, yanlu\}@microsoft.com  \\
}

\begin{document}

    \maketitle

    \begin{abstract}
        Most of the existing neural video compression methods adopt the predictive coding framework, which first generates the predicted frame and then encodes its residue with the current frame. However, as for compression ratio, predictive coding is only a sub-optimal solution as it uses simple subtraction operation to remove the redundancy across frames.
        In this paper, we propose a deep contextual video compression framework to enable a paradigm shift from  predictive coding to  conditional coding. In particular, we try to answer the following questions: how to define, use, and learn condition under a deep video compression framework. To  tap the potential of conditional coding, we propose using feature domain context as condition. This enables us to  leverage the high dimension context to carry rich information to both the encoder and the decoder, which helps reconstruct the high-frequency contents for higher video quality. Our framework is also extensible, in which the condition can be flexibly designed.
        Experiments show that our method can significantly outperform the previous state-of-the-art (SOTA) deep video compression methods. When compared with x265 using \textit{veryslow} preset, we can achieve 26.0\% bitrate saving for 1080P standard test videos.

    \end{abstract}

    \section{Introduction}

    From H.261 \cite{girod1995comparison} developed in 1988 to the just released H.266 \cite{bross2021developments} in 2020, all traditional video coding standards are based on a predictive coding paradigm, where the  predicted frame is first generated by handcrafted modules and then the residue between the current frame and the predicted frame is encoded and decoded. Recently, many deep learning (DL)-based video compression methods \cite{lu2019dvc,lu2020end,lu2020content,Yang_2020_CVPR,lin2020m,Djelouah_2019_ICCV, agustsson2020scale, wu2018vcii, yang2020hierarchical} also adopt the predictive coding framework to encode the residue, where all handcrafted modules  are merely replaced by neural networks.

    Encoding residue is a simple yet efficient manner for video compression, considering the strong temporal correlations among frames. However, residue coding is not optimal to encode the current frame $x_{t}$ given the predicted frame $\tilde{x}_{t}$, because it only uses handcrafted subtraction operation to remove the redundancy across frames. The entropy of residue coding is greater than or equal to that of conditional coding  \cite{ladune2020optical}:
    $H(x_{t}-\tilde{x}_{t})\geq H(x_{t}|\tilde{x}_{t})$, where $H$ represents the Shannon entropy.
    Theoretically, one pixel in frame $x_{t}$ correlates to all the pixels in the previous decoded frames and the pixels already been decoded in  $x_{t}$.
    For traditional video codec, it is impossible to use the handcrafted rules to explicitly explore the correlation by taking all of them into consideration due to the huge space. Thus, residue coding is widely adopted as a special extremely simplified case of conditional coding, with the very strong assumption that the current pixel only has the correlation with the predicted pixel.
    DL opens the door to automatically explore correlations in a huge space. Considering the  success of DL in image compression \cite{balle2018variational, minnen2018joint}, which just uses autoencoder to explore correlation in image, why not use network to build the conditional coding-based autoencoder to explore correlation in video
    rather than restricting our vision into residue coding?

    \begin{figure}
        \centering
        \includegraphics[width=400pt, height=125pt]{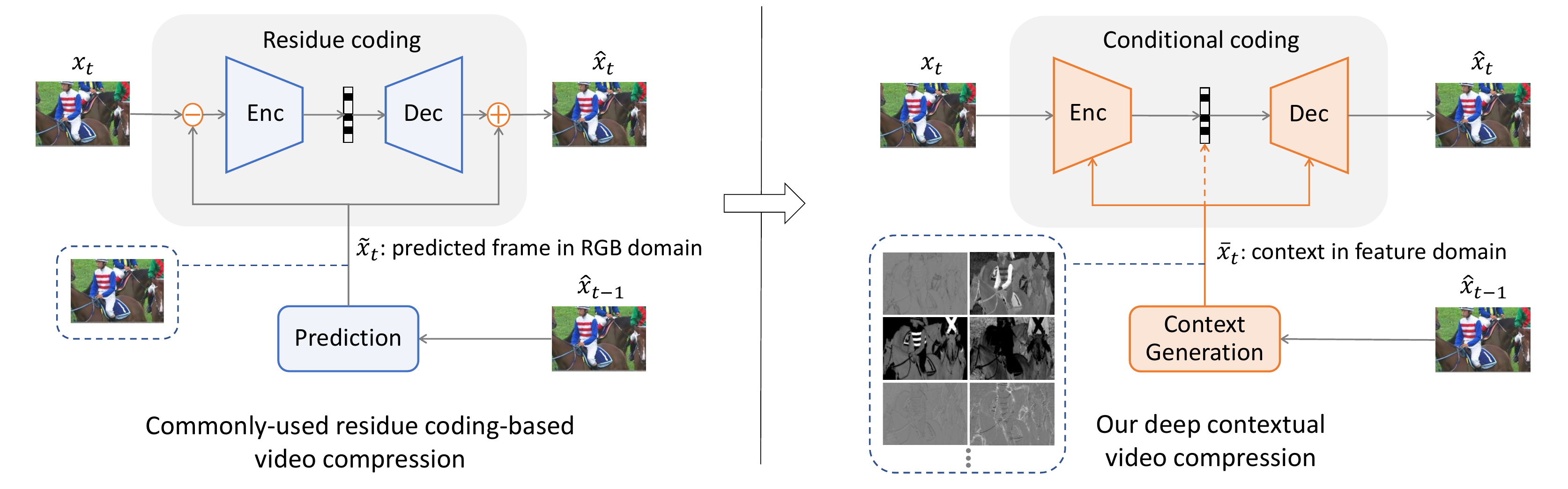}\\

        \caption{ Paradigm shift from residue coding-based framework to conditional coding-based  framework. $x_{t}$  is the current frame.  $\hat{x}_{t}$ and $\hat{x}_{t-1}$ are the current and previous decoded frames. The orange dashed line means that the context is also used for entropy modeling.
        }\label{insight}
    \end{figure}
    When we design the conditional coding-based solution, a series of questions naturally come up: \textit{What is condition? How to use condition? And how to learn condition?}
    Technically speaking, condition can be anything that may be helpful to compress the current frame.
    The  predicted frame can be used as condition but it is not necessary to restrict it as  the only representation of condition.
    Thus, we define the condition as learnable contextual features with arbitrary dimensions.
    Along this idea, we propose a deep contextual video compression (DCVC) framework to utilize condition in a unified, simple, yet efficient approach. The diagram of our DCVC framework is shown in Fig. \ref{insight}.
    The contextual information is used as part of the input of contextual encoder, contextual decoder, as well as the entropy model. In particular,  benefiting from the temporal prior provided by context, the entropy model itself is temporally adaptive, resulting in a richer and more accurate model.
    As for how to learn condition,  we propose using  motion estimation and motion compensation (MEMC) at feature domain.  The MEMC can guide the model where to extract useful context.
    Experimental results demonstrate the effectiveness of the proposed DCVC. For 1080p standard test videos, our DCVC can achieve 26.0\% bitrate saving over x265 using \textit{veryslow} preset, and 16.4\% bitrate saving over previous SOTA DL-based model DVCPro \cite{lu2020end}.

    Actually, the concept of conditional coding has appeared in  \cite{liu2020conditional,rippel2019learned,ladune2020optical, ladune2021conditional}. However, these works are only designed for partial module (e.g., only entropy model or encoder) or need handcrafted operations to filter which content should be conditionally coded.  By contrast, our framework is a more comprehensive solution which considers all of encoding, decoding, and entropy modeling. In addition,  the proposed DCVC is an extensible conditional coding-based framework, where the condition can be flexibly designed. Although this paper proposes using feature domain MEMC to generate contextual features and demonstrates its effectiveness,
    we still think it is an  open question worth further investigation for higher compression ratio.

    Our main contributions are four-folded:
    \begin{itemize}
        \item We design a deep contextual video compression framework based on conditional coding. The definition, usage, and   learning manner of condition are all innovative.  Our method can achieve higher compression ratio than previous residue coding-based  methods.
        \item We propose a simple yet efficient approach using context to help the encoding, decoding, as well as the entropy modeling. For entropy modeling, we design a model which utilizes spatial-temporal correlation for higher compression ratio or only utilizes temporal correlation for fast speed.
        \item We define the condition as the context in feature domain. The context with higher  dimensions can provide richer information to help reconstruct the high frequency contents.
        \item  Our framework is extensible. There exists great  potential in boosting compression ratio by better defining, using, and learning   the condition.
    \end{itemize}
    \section{Related works}

    \paragraph{Deep image compression}
    Recently there are many works for deep image compression. For example, the compressive autoencoder \cite{theis2017lossy} could get comparable results with  JPEG 2000. Subsequently, many works boost the performance by more advanced entropy models and network structures. For example,  Ball{\'e} \textit{ et al.} proposed the factorized \cite{ball2017endtoend} and hyper prior \cite{balle2018variational} entropy models.  The method based on hyper prior catches up H.265 intra coding. The entropy model jointly utilizing hyper prior and auto regressive context outperforms  H.265 intra coding.
    The method with Gaussian mixture model \cite{cheng2020learned}   is comparable with H.266 intra coding. 
    For the network structure, some RNN (recurrent neural network)-based methods \cite{toderici2015variable, toderici2017full, johnston2018improved} were proposed in the early development stage, but most of recent methods are based on CNN (convolutional  neural network).

    \paragraph{Deep video compression}
    Existing works for deep video compression can be classified into two categories, i.e.
    non-delay-constrained and delay-constrained. For the first category, there is no restriction on reference frame location, which means that the reference frame can be from future. For example, Wu \textit{ et al.} \cite{wu2018vcii} proposed  interpolating the predicted frame from previous and future frames, and then  the frame residue is encoded.  Djelouah  \textit{et al.}\cite{Djelouah_2019_ICCV} also followed this coding structure and introduced the optical flow estimation network to get better predicted frame.
    Yang \textit{ et al.} \cite{Yang_2020_CVPR} designed a  recurrent enhancement module for this coding structure.  In addition, 3D autoencoder was proposed   to encode group of pictures in  \cite{pessoa2020end, habibian2019video}. This is a natural extension of  deep image compression by increasing the dimension of input. It is noted that  this coding manner will bring larger  delay and the GPU memory cost will be significantly increased.   For the delay-constrained methods, the reference frame is only from the previous frames. For example, Lu \textit{ et al.} \cite{lu2019dvc} designed the DVC model, where all modules in traditional hybrid video codec are  replaced by networks.  Then the improved model DVCPro which adopts more advanced entropy model from \cite{minnen2018joint} and deeper network was proposed in \cite{lu2020end}.  Following the similar framework with DVC,  Agustsson \textit{ et al.} designed a more advanced optical flow estimation  in scale space. Hu \textit{ et al.} \cite{hu2020improving} considered the rate distortion optimization when encoding  motion vector (MV). 
    In \cite{lin2020m}, the single reference frame is extended to multiple reference frames.
    Recently, Yang \textit{ et al.} \cite{Yang_2020_CVPR} proposed an RNN-based MV/residue encoder and decoder. In \cite{yang2020hierarchical}, the residue is adaptively scaled by learned parameter.

    Our research belongs to the delay-constrained method as it can be applied in more  scenarios, e.g. real time communication. Different from the above works, we design a conditional coding-based framework rather than following the  commonly-used residue coding. Other video tasks show that utilizing temporal information as condition is helpful \cite{bao2019depth, niklaus2020softmax}. For video compression, recent works in \cite{liu2020conditional}, \cite{rippel2019learned}, and \cite{ladune2020optical, ladune2021conditional} have some investigations on condition coding.  In \cite{liu2020conditional}, the conditional coding is only designed for entropy modeling. However, due to the lack of MEMC, the compression ratio is not high, and the method  in  \cite{liu2020conditional} cannot outperform  DVC in terms of PSNR.
    By contrast,  our  conditional coding designed for encoding, decoding, and entropy modeling can significantly outperform  DVCPro. In \cite{rippel2019learned}, only encoder takes the conditional coding. However, for decoder, the residual coding is still adopted. As a latent state is used, the framework in \cite{rippel2019learned} is difficult to train \cite{lin2020m}. By contrast, we use explicit MEMC to guide the context learning, which is easier to train.
    In \cite{ladune2020optical, ladune2021conditional}, the video contents need to be explicitly classified into skip and non-skip modes, where only the contents with non-skip mode   use the conditional coding.
    By contrast,  our method does not need the handcrafted operation to decompose the video. In addition, the condition in DCVC is the context in feature domain, which  has much larger capacity. In summary, when compared with \cite{liu2020conditional}, \cite{rippel2019learned}, and \cite{ladune2020optical, ladune2021conditional}, the definition, usage, and  learning manner of the condition in DCVC are all innovative.

    \section{Proposed method}
    In this section, we present the details of the proposed  DCVC.  We first describe the whole framework of DCVC.
    Then we introduce the entropy model for compressing the latent codes, followed by the approach of learning the context.
    At last, we  provide the details about training.

    \begin{figure*}
        \centering
        \includegraphics[width=380pt, height=170pt]{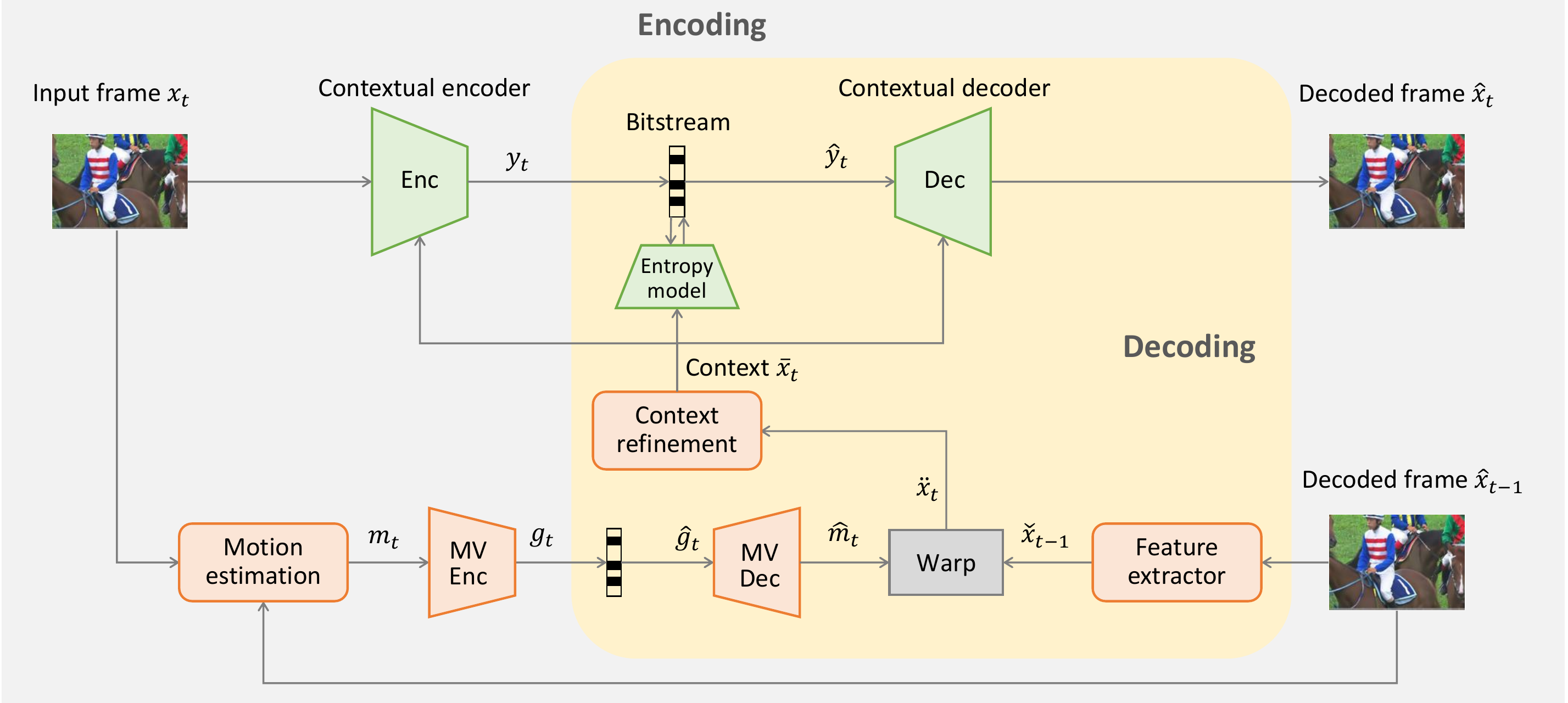}\\

        \caption{ The framework of our DCVC.
        }\label{framework}
    \end{figure*}

    \subsection{The framework of DCVC}
    In traditional  video codec, the inter frame coding  adopts the  residue coding, formulated as:
    \begin{equation}\label{hybrid_residue_coding}
        \hat{x}_{t} = f_{dec}\big(\lfloor f_{enc}(x_{t}- \tilde{x}_{t} )\rceil \big) + \tilde{x}_{t} \; with \; \tilde{x}_{t} = f_{predict}(\hat{x}_{t-1}).
    \end{equation}
    $x_{t}$  is the current frame.  $\hat{x}_{t}$ and $\hat{x}_{t-1}$ are the  current and previous decoded frames. For simplification, we use single reference frame in the formulation.
    $f_{enc}(\cdot)$ and $f_{dec}(\cdot)$ are the residue  encoder and decoder. $\lfloor \cdot \rceil$ is the quantization operation.
    $f_{predict}(\cdot)$ represents the  function for generating the predicted frame  $\tilde{x}_{t}$.
    In traditional video codec, $f_{predict}(\cdot)$ is implemented in the manner of  MEMC, which uses handcrafted coding tools to search the best  MV and then interpolates the predicted frame. For most existing DL-based video codecs \cite{lu2019dvc,lu2020end,lu2020content,Yang_2020_CVPR,lin2020m,Djelouah_2019_ICCV, agustsson2020scale}, $f_{predict}(\cdot)$ is the MEMC which is totally composed of neural networks.

    In this paper, we do not adopt the commonly-used residue coding  but try to design a conditional coding-based framework for higher compression ratio. Actually, one straightforward conditional coding manner is directly using the predicted frame  $\tilde{x}_{t}$ as the condition:
    \begin{equation}\label{hybrid_residue_coding}
        \hat{x}_{t} = f_{dec}\big(\lfloor f_{enc}(x_{t} | \tilde{x}_{t} )\rceil \;| \;\tilde{x}_{t} \big)  \; with \; \tilde{x}_{t} = f_{predict}(\hat{x}_{t-1}).
    \end{equation}
    However, the condition is still restricted  in pixel domain with low channel dimensions. 
    This will limit the  model capacity.  Now that the conditional coding is used, why not let model learn the condition by itself? Thus, this paper proposes a  contextual video compression  framework, where we use network to generate context rather than the predicted frame. Our framework can be formulated as:
    \begin{equation}\label{hybrid_residue_coding}
        \hat{x}_{t} = f_{dec}\big(\lfloor f_{enc}(x_{t} | \bar{x}_{t} )\rceil \;| \;\bar{x}_{t} \big)  \; with \; \bar{x}_{t} = f_{context}(\hat{x}_{t-1}).
    \end{equation}
    $f_{context}(\cdot)$ represents the function of generating context $\bar{x}_{t}$.  $f_{enc}(\cdot)$ and $f_{dec}(\cdot)$ are the contextual  encoder and decoder, which are different from residue encoder and decoder.     Our DCVC framework is illustrated in Fig. \ref{framework}.

\begin{figure*}
        \centering
        \includegraphics[width=400pt, height=240pt]{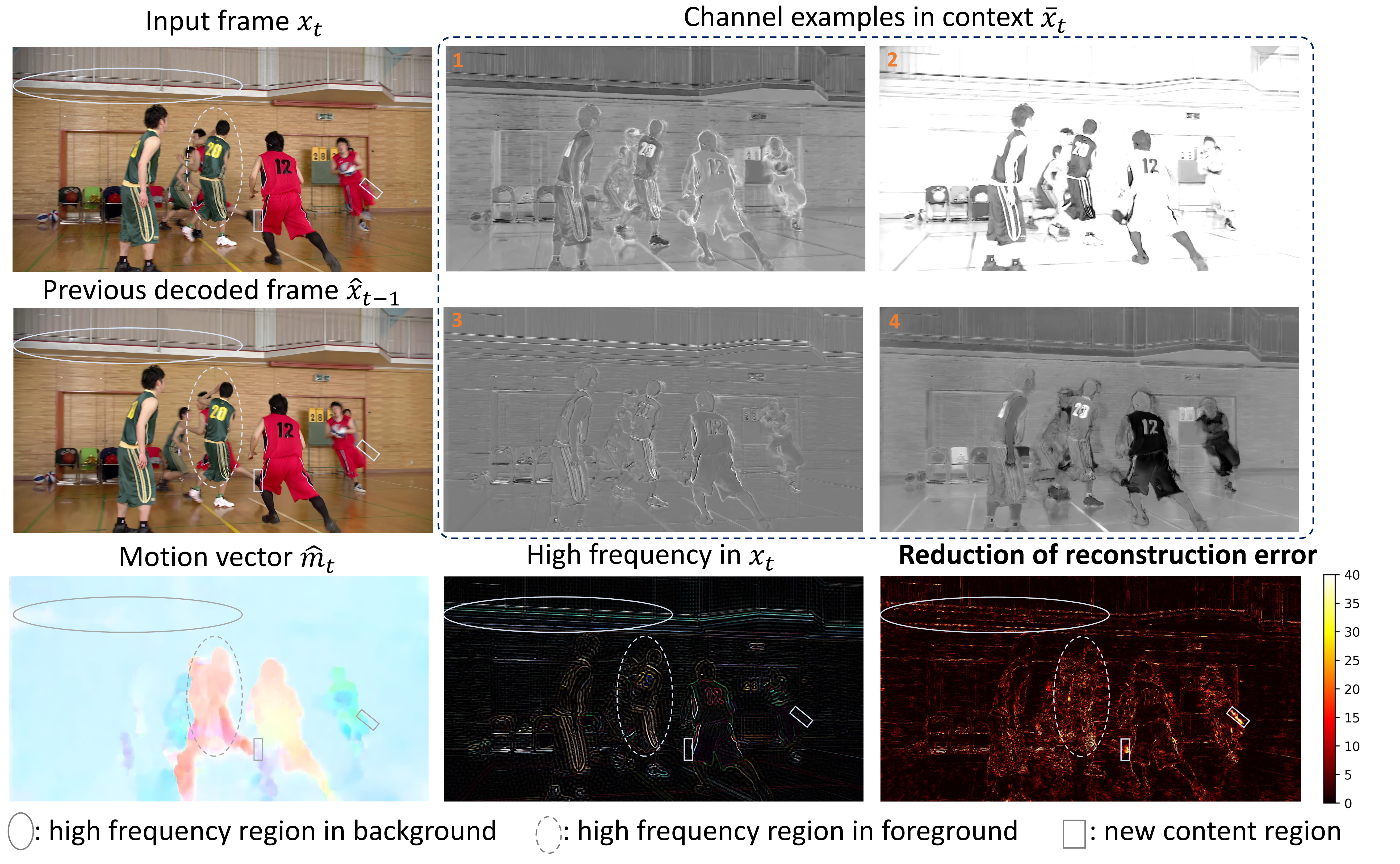}\\

        \caption{ Visual examples from \textit{videoSRC05} in  MCL-JCV\cite{wang2016mcl} dataset.  The high frequency in $x_{t}$ is decomposed by discrete cosine transform. It shows that DCVC improves the reconstruction of high frequency contents in both background with small motion and foreground with large motion. In addition, DCVC is good at encoding the new content region caused by motion, where the reconstruction error can be significantly decreased compared with residue coding-based framework DVCPro\cite{lu2020end}. The BPP (bits per pixel)  of DCVC (0.0306)  is smaller than that of DVCPro  (0.0359).
        }\label{prediction_comparision}
    \end{figure*}

    To provide  richer and more correlated condition for encoding  $x_{t}$, the context is in the feature domain with higher dimensions.
    In addition, due to  the large capacity of context, different channels therein have the  freedom to extract different kinds of  information.   Here we give an analysis example in Fig. \ref{prediction_comparision}.  In the figure, the upper right part shows  four channel examples in context. Looking into the  four channels, we can find   different channels have different focuses.  For example, the third channel seems to put a lot of emphases on   the high frequency contents when compared with the visualization of high frequency in  $x_{t}$.  By contrast, the second and fourth channels look like extracting color information, where the second channel focuses on green color and the fourth channel emphasizes the red color.
    Benefiting from these various contextual features, our DCVC can achieve better reconstruction quality, especially for the complex textures with lots of high frequencies. The bottom right image in Fig. \ref{prediction_comparision} shows the reconstruction error reduction of DCVC when compared with residue coding-based framework. From this comparison, DCVC can achieve non-trivial error decrease on the  high frequency regions in both background and foreground, which are hard to compress for many video codecs.

    As shown in Fig. \ref{framework}, the encoding and decoding of the current frame are both conditioned on the context $\bar{x}_{t}$. Through contextual encoder, ${x}_{t}$ is  encoded  into  latent codes $y_{t}$.  $y_{t}$ is then quantized  as $\hat{y}_{t}$ via rounding operation. Via the contextual decoder, the reconstructed frame $\hat{x}_{t}$ is  finally obtained.  In our design, we use network to automatically  learn the correlation between ${x}_{t}$  and   $\bar{x}_{t}$ and then remove the redundancy rather than using fixed subtraction operation in  residue coding.
    From another perspective, our method also has the ability to adaptively use the context.  Due to the motion in video,  new contents often appear in the object boundary regions.
    These new contents probably cannot find a good reference in previous decoded frame.
    In this situation, the DL-based video codec with  frame residue coding is still forced to encode the residue.
    For the new contents, the residue can be very large  and the inter coding via subtraction operation may be worse than the intra  coding. By contrast, our conditional coding framework has the capacity to adaptively  utilize the condition. For the new contents, the model can adaptively tend to learn intra coding. As shown in reconstruction error reduction in Fig. \ref{prediction_comparision}, the reconstruction error of new contents can be significantly reduced.

    In addition, this paper  not only proposes using the context $\bar{x}_{t}$  to generate the latent codes, but also proposes utilizing it to build the entropy model. More details are introduced in the next subsection.

    \subsection{Entropy model}

\begin{figure}
        \centering
        \includegraphics[width=260pt, height=120pt]{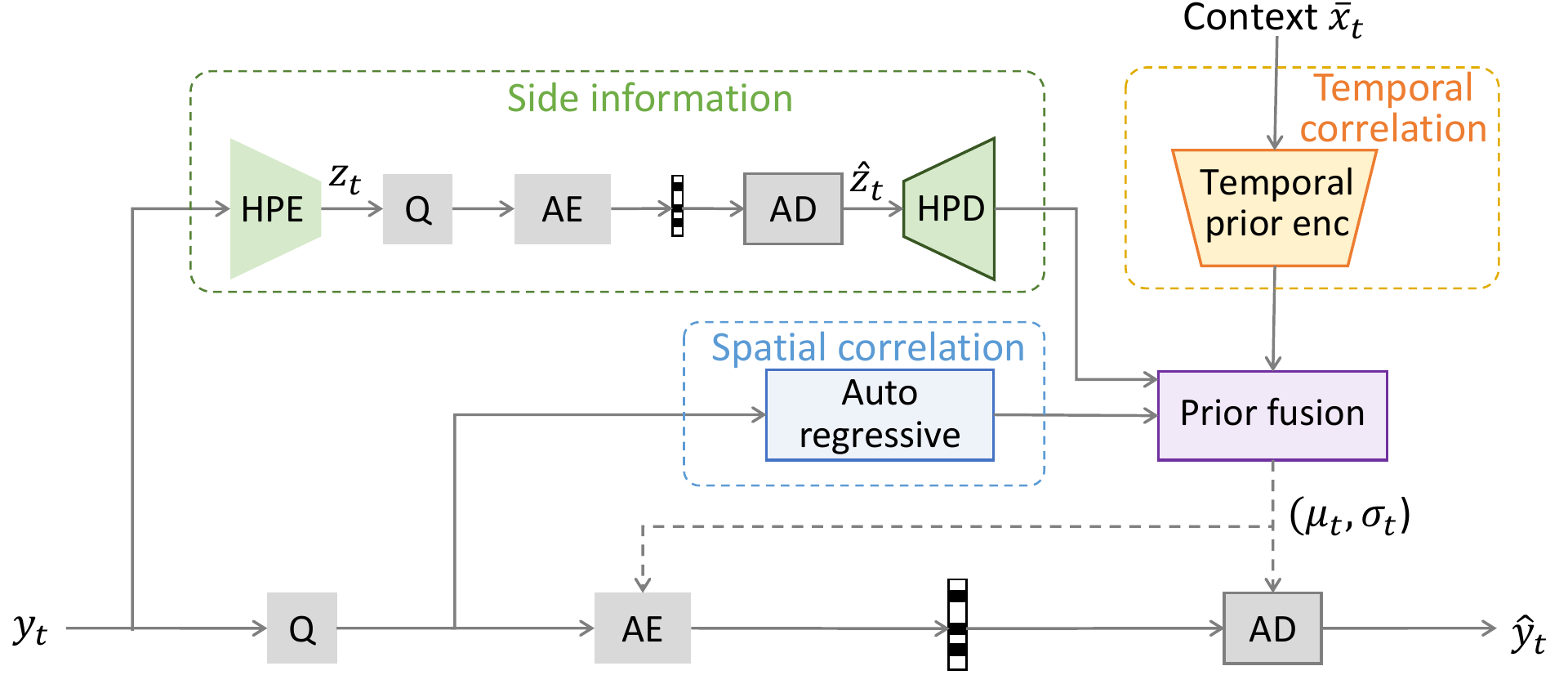}\\

        \caption{Our entropy model used to encode the quantized latent codes $\hat{y}_{t}$. HPE and HPD are hyper prior encoder and decoder. Q means quantization. AE and AD are arithmetic encoder and decoder.
        }\label{entropy_model}
    \end{figure}

    According to \cite{shannon2001mathematical},   the cross-entropy between the estimated probability distribution and the actual latent code
    distribution is a tight lower bound of the actual bitrate, namely
    \begin{equation}\label{cross_entroy}
        R(\hat{y}_{t}) \geq  \mathbb{E}_{\hat{y}_{t} \sim q_{\hat{y}_{t}}}[-log_{2}  p_{\hat{y}_{t}}(\hat{y}_{t})],
    \end{equation}
    $p_{\hat{y}_{t}}(\hat{y}_{t})$ and $q_{\hat{y}_{t}}(\hat{y}_{t})$ are estimated  and  true probability mass functions of quantized latent codes $\hat{y}_{t}$, respectively.
    Actually, the arithmetic coding almost can encode the latent codes at the bitrate of cross-entropy. The gap between actual bitrate $R(\hat{y}_{t})$  and  the bitrate of cross-entropy is  negligible.  So our target is  designing an  entropy model which can accurately estimate the probability distribution of latent codes $p_{\hat{y}_{t}}(\hat{y}_{t})$.
    The framework of our entropy model  is illustrated in Fig. \ref{entropy_model}.
    First, we use the hyper prior model \cite{balle2018variational} to learn the hierarchical prior and use auto regressive network \cite{minnen2018joint} to learn the spatial prior.
    The two priors (hierarchical prior and spatial prior) are commonly-used in deep image compression. However, for video, the latent codes also have the temporal correlation. Thus, we propose using the context $\bar{x}_{t}$ to generate the temporal prior. As shown in Fig. \ref{entropy_model}, we design a temporal prior encoder  to explore the temporal correlation. The prior fusion network will learn to fuse the three different priors and then estimate the mean and scale of  latent code distribution. In this paper, we follow the existing work \cite{PyTorchVideoCompression} and assume that $p_{\hat{y}_{t}}(\hat{y}_{t})$ follows the Laplace distribution. The formulation of  $p_{\hat{y}_{t}}(\hat{y}_{t})$ is:
    \begin{equation}\label{distribution}
        \begin{split}
            p_{\hat{y}_{t}}(\hat{y}_{t}|\bar{x}_{t}, \hat{z}_{t}) = \prod_{i}\big(\mathcal{L}(\mu_{t,i},\sigma_{t,i}^{2})\ast  \mathcal{U}(-\frac{1}{2},\frac{1}{2})\big) (\hat{y}_{t,i})
            \;\; \\with\;\;  \mu_{t,i},\sigma_{t,i} = f_{pf}\big(f_{hpd}(\hat{z}_{t}), f_{ar}(\hat{y}_{t,<i})  ,f_{tpe}(\bar{x}_{t})\big).
        \end{split}
    \end{equation}
    The index $i$ represents the spatial location.
    $f_{hpd}(\cdot)$ is   the hyper prior decoder network. $f_{ar}(\cdot)$ is the auto regressive network.
    $f_{tpe}(\cdot)$ is the  specially designed  temporal prior encoder network.
    $f_{pf}(\cdot)$ denotes the prior fusion network.
    In our entropy model,  $f_{ar}(\hat{y}_{t,<i})$ and $f_{tpe}(\bar{x}_{t})$ provide the spatial and temporal priors, respectively. $f_{hpd}(\hat{z}_{t})$ provides the supplemental side information which cannot be learned from  spatial and temporal correlation.

    The entropy model formulated in Eq. \ref{distribution} utilizes spatial prior for higher compression ratio. However, the operations about spatial prior are non-parallel and then lead to slow encoding/decoding speed. By contrast, all operations about the proposed temporal prior are parallel.  Thus, we also provide a solution which removes spatial prior but relies on temporal prior for acceleration, namely $\mu_{t,i},\sigma_{t,i} = f_{pf}\big(f_{hpd}(\hat{z}_{t}),f_{tpe}(\bar{x}_{t})\big)$.

    \subsection{Context learning}
    For how to learn the  context, one alternative solution is directly using a plain network  composed by several convolutional layers, where the input is the previous decoded frame $\hat{x}_{t-1}$ and the output is  $\bar{x}_{t}$.  However, it is hard for a plain network to extract useful information without supervision.
    Video often contains various contents and there  exist a lot of complex motions. For a position in the current frame, the collocated position in the reference frame may have less correlation. In this situation, the collocated position in context  $\bar{x}_{t}$ is also less correlated to the position in $x_{t}$, and the less correlated context cannot facilitate the compression of  $x_{t}$.
    For this reason, we also adopt the idea  of MEMC.
    But different from commonly usage of applying MEMC in pixel domain, we propose performing MEMC in feature domain. The context generation function $f_{context}$ is designed as:
    \begin{equation}\label{warp}
        f_{context}(\hat{x}_{t-1}) = f_{cr}\big (warp(f_{fe}(\hat{x}_{t-1}),\hat{m}_{t})\big )
    \end{equation}
    We first design a feature extraction network $\check{x}_{t-1}= f_{fe}(\hat{x}_{t-1})$ to convert the reference frame from pixel domain to feature domain. At the same time, we  use the optical flow estimation network \cite{ranjan2017optical} to learn the  MV between the  reference frame $\hat{x}_{t-1}$ and the current  frame ${x}_{t}$.
    The  MV ${m}_{t}$  is then encoded and decoded. The decoded $\hat{m}_{t}$  guides the network where to extract the context through warping operation $\ddot{x}_{t} = warp(\check{x}_{t-1}, \hat{m}_{t})$.
    The  $\ddot{x}_{t}$  is kind of relatively rough context  because the warping operation may introduce some spatial discontinuity. Thus, we design a context refinement  network   $f_{cr}(\cdot)$ to obtain the final context $\bar{x}_{t}= f_{cr} (\ddot{x}_{t})$.
    In function $ f_{context}(\cdot)$, MV $\hat{m}_{t}$ not only guides the network where to extract the context but also  enables network to learn context from a larger reference region when compared with the solution without MEMC.


    \subsection{Training}
    \label{Training}
    The target of video compression is using least bitrate  to  get the best reconstruction quality. Thus, the training loss consists of two metrics:
    \begin{equation}\label{RDO}
        L = \lambda \cdot D + R 
    \end{equation}
    $\lambda$ controls the trade-off between the distortion $D$ and the bitrate cost $R$.
    $D$ can be MSE (mean squared error) or MS-SSIM (multiscale structural similarity) for different targets. During the training, $R$ is  calculated as  the  cross-entropy between the true and estimated probability of the  latent codes.


    For the learning rate, it is set as 1e-4  at the start and  1e-5  at the fine-tuning stage. The training batch size is set as 4. 
    For comparing DCVC with other methods, we follow \cite{lu2020content} and train 4 models with different  $\lambda$ s \{MSE: 256, 512, 1024, 2048; MS-SSIM: 8, 16, 32, 64\}.

    \section{Experimental results}\label{Sec_Experimental_Results}

    \subsection{Experimental settings}
    \label{Experimental_setting}
    \textbf{Training data} We use the training part in Vimeo-90k septuplet  dataset \cite{xue2019video} as our training data. 
     During the training, we will randomly crop videos into 256x256 patches. 

    \textbf{Testing data} The testing data includes HEVC Class B (1080P), C (480P), D (240P), E (720P) from the common test conditions \cite{bossen2013common} used by codec  standard community.  In addition, The 1080p videos from MCL-JCV\cite{wang2016mcl} and UVG\cite{uvg} datasets  are also tested.

    \textbf{Testing settings} The GOP (group of pictures) size is same with \cite{lu2020end}, namely 10 for HEVC videos and 12 for non-HEVC videos. As this paper only focuses on inter frame coding, for intra frame coding, we directly use  existing deep image compression models  provided by CompressAI \cite{begaint2020compressai}.  We use \textit{ cheng2020-anchor} \cite{cheng2020learned} for MSE target and use \textit{hyperprior} \cite{balle2018variational} for MS-SSIM target.

    According to the performance comparison in \cite{vcip_XuLYT20, TutorialVCIP, PyTorchVideoCompression},  DVCPro\cite{lu2020end}  is one previous SOTA DL-based codec among recent works \cite{Djelouah_2019_ICCV,Yang_2020_CVPR,agustsson2020scale, lu2020content, hu2020improving}.
    Thus, we compare DVCPro in our paper. In addition, its predecessor DVC\cite{lu2019dvc} is also tested.  It is noted that, for fair comparison, DVC and DVCPro are retested using the same intra frame coding with DCVC. 
    For traditional  codecs,  x264 and x265 encoders \cite{FFMPEG} are tested. The settings of  these two encoders are same with \cite{lu2020end} except two options. One is  that we use the \textit{veryslow} preset  rather than \textit{veryfast} preset.   \textit{Veryslow} preset can achieve higher compression ratio   than \textit{veryfast} preset. Another is we use the constant quantization parameter setting rather than constant  rate factor  setting to avoid the influence of rate control.

    \begin{table}[]
        \centering
        \caption{The  BD-Bitrate comparison}
        \label{bd_result}
        {\def\arraystretch{1.2}
            \begin{adjustbox}{width=1\textwidth}
                \begin{tabular}{lcccccc}
                    \toprule
                    Method & MCL-JCV & UVG & HEVC Class B & HEVC Class C & HEVC Class D & HEVC Class E \\ \hline
                    \textbf{DCVC (proposed) }& \textbf{-23.9\%} & \textbf{-25.3\%} &   \textbf{-26.0\%} &   \textbf{-5.8\%} &  \textbf{-17.5\%} &  \textbf{-11.9\% }   \\
                    DVCPro \cite{lu2020end}         &  -4.1\% &  -7.9\% &   -9.0\%  &   7.2\%  &  -6.9\%  &  17.2\%     \\
                    x265 (\textit{veryslow})          &  0.0\% &  0.0\% &   0.0\%  &   0.0\%  &  0.0\%  &  0.0\%     \\
                    DVC \cite{lu2019dvc}            &  13.3\% &  17.2\% &   7.9\%   &   15.1\% &  7.2\%   &  21.1\%     \\
                    x264 (\textit{veryslow}) &  32.7\% &  30.3\% &   35.0\%  &   19.9\% &  15.5\%  &  50.0\%     \\
                    \bottomrule
                \end{tabular}
            \end{adjustbox}
        }
        \begin{flushleft}
            The anchor is x265 (\textit{veryslow}). Negative number means bitrate saving and positive number means bitrate increase.
        \end{flushleft}
    \end{table}

    \begin{figure*}
        \centering
        \includegraphics[width=400pt, height=220pt]{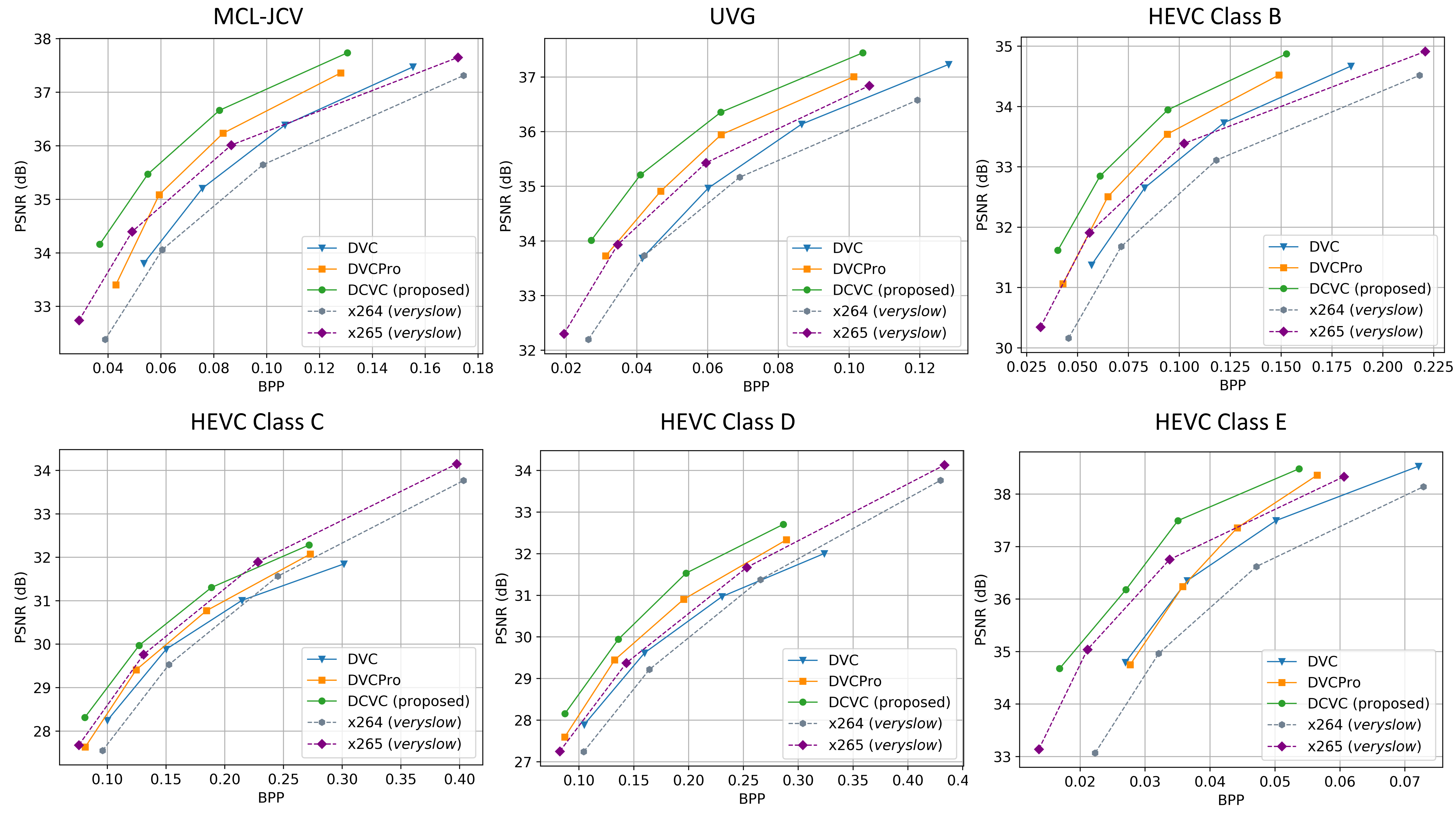}\\

        \caption{ PSNR and bitrate comparison.  The horizontal axis is  bits per pixel (BPP) representing bitrate cost and vertical axis is PSNR representing reconstruction quality.
        }\label{rd_curve_psnr}
    \end{figure*}

    \subsection{Performance comparison}

\begin{figure*}
        \centering
        \includegraphics[width=400pt, height=220pt]{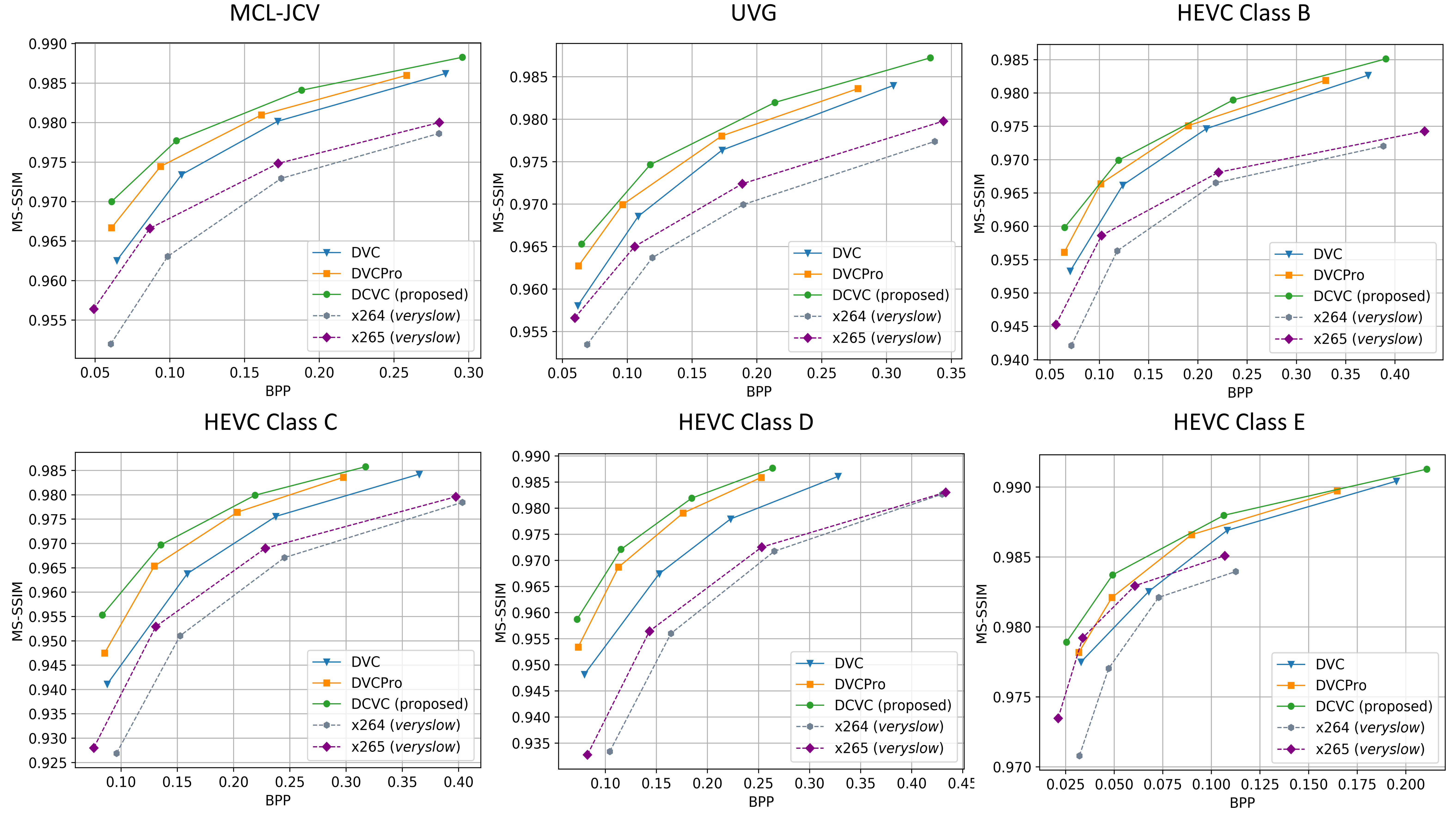}\\

        \caption{ MS-SSIM and bitrate comparison. The DL-based codecs are fine-tuned for MS-SSIM.
        }\label{rd_curve_msssim}
    \end{figure*}

    \textbf{Compression ratio} Fig. \ref{rd_curve_psnr} and Fig. \ref{rd_curve_msssim}  show the rate-distortion curves among these methods, where the distortion in Fig. \ref{rd_curve_psnr} is measured by PSNR and the distortion in Fig. \ref{rd_curve_msssim} is measured by MS-SSIM. From these figures, we can find that our DCVC model can outperform DVCPro for all bitrate ranges.   Table \ref{bd_result} gives the corresponding BD-Bitrate \cite{bjontegaard2001calculation} results in terms of PSNR.
    When compared with x265 using \textit{veryslow} preset,  DVCPro achieves 4.1\%, 7.9\%, 9.0\%,  and 6.9\% bitrate saving on MCL-JCV, UVG, HEVC Class B, and D, respectively. However, for HEVC Class C and E, DVCPro  performs worse, and there are 7.2\% and 17.2\% bitrate increases.  By contrast,  our DCVC can outperform x265 on all datasets. For three 1080p datasets (MCL-JCV, UVG, HEVC Class B), the bitrate savings are  23.9\%, 25.3\%, and 26.0\%, respectively. For low resolution videos HEVC Class C  and D, the improvements are also adequate, i.e. 5.8\% and 17.5\%. For HEVC Class E with relatively  small motion,  the bitrate saving is 11.9\%. From these comparisons, we can find that our DCVC can significantly outperform DVCPro and x265 for various videos with different resolutions and different content characteristics.

    In addition, we can find that DCVC can achieve larger improvement on high resolution videos. This is because that high resolution video contains more textures with high frequency. For this kind of video, the feature domain  context with higher dimensions is more helpful and able to carry richer contextual information to reconstruct the high frequency contents.

    \textbf{Complexity} The MACs (multiply–accumulate operations) are 2268G for DCVC and 2014G for DVCPro, and there is about 13\% increase. However, the actual inference time per 1080P frame is 857 ms for DCVC and 849 ms for DVCPro on P40 GPU, and there is only about 1\% increase, mainly due to the parallel ability of GPU.

    \subsection{Ablation study}
    \textbf{Conditional coding and temporal prior}  In our DCVC, we propose using concatenation-based conditional coding to replace subtraction-based residue coding. At the same time, we design the temporal prior for entropy model. To verify the  effectiveness of these ideas, we make the ablation study shown in Table \ref{study_conditionalcoding}, where the baseline is our final solution (i.e. temporal prior + concatenating context feature). From this table, we can find that both concatenating RGB prediction and concatenating context feature improve the compression ratio. It verifies the benefit of conditional coding compared with residue coding. In addition, we can find that the improvement of concatenating context feature is much larger than that of concatenating RGB prediction. It shows the advantage of context in feature domain.
     From Table \ref{study_conditionalcoding}, we also find that the proposed temporal prior further boosts the performance, and its improvement under conditional coding (no matter concatenating RGB prediction or concatenating context feature) is larger than that under residue coding.  These results demonstrate the advantage of our  ideas.

      \begin{table}[]
        \centering
        \caption{Ablation study on conditional coding and temporal prior}
        \label{study_conditionalcoding}
        {\def\arraystretch{1.2}
            \begin{adjustbox}{width=0.9\textwidth}
                \begin{tabular}{cccc}
                    \toprule
                    Temporal prior & Concatenate context feature & Concatenate RGB prediction & Bitrate increase  \\ \hline
                    $\checkmark$        &$\checkmark $  &                                 &    0.0\%     \\
                   $ \checkmark$        &                           & $ \checkmark$       &   5.4\%     \\
                                                    & $\checkmark$  &                                 &    4.6\%     \\
                                                    &                            &  $\checkmark$      &    8.7\%     \\
                   $ \checkmark$        &                            &                                &    11.2\%     \\
                                                    &                            &                                 &    12.9\%     \\
                    \bottomrule
                \end{tabular}
            \end{adjustbox}
        }
    \end{table}

         \textbf{Entropy model}  In DCVC, besides the hyper prior model, the entropy model for compressing the quantized latent codes $\hat{y}_{t}$ utilizes both spatial and temporal priors  for higher compression ratio.  However, the drawback of spatial prior is slow encoding/decoding speed as it brings spatial dependency and is non-parallel. By contrast, all operations about the proposed temporal prior are parallel.  Thus, our DCVC also supports removing spatial prior but relies on temporal prior for acceleration. Benefiting from the rich temporal context, the model without spatial prior only has small bitrate increase.
         Table  \ref{study_entropymodel} compares the performance influence of spatial and temporal priors.  From this table, we can find that the performance has large drop if both priors are disabled.  When enabling either of these two priors, the performance can be improved a lot. When enabling both of them, the performance can be further improved. However, considering the trade-off between complexity and compression ratio, the solution only using hyper prior and temporal prior is better.  These results show the advantage of our temporal prior-based entropy model.

         \begin{table}[]
         \centering
         \caption{Ablation study on entropy model}
         \label{study_entropymodel}
         {\def\arraystretch{1.2}
             \begin{adjustbox}{width=0.65\textwidth}
                 \begin{tabular}{lc}
                     \toprule
                     Entropy model &   Bitrate increase  \\ \hline
                     hyper prior + spatial prior + temporal prior       &    0.0\%     \\
                     hyper prior +  temporal prior                            &    3.8\%     \\
                     hyper prior + spatial prior                                   &    4.6\%     \\
                     hyper prior                                                         &    60.9\%     \\
                     \bottomrule
                 \end{tabular}
             \end{adjustbox}
         }
     \end{table}

 \begin{figure}
     \centering
     \includegraphics[width=220pt, height=180pt]{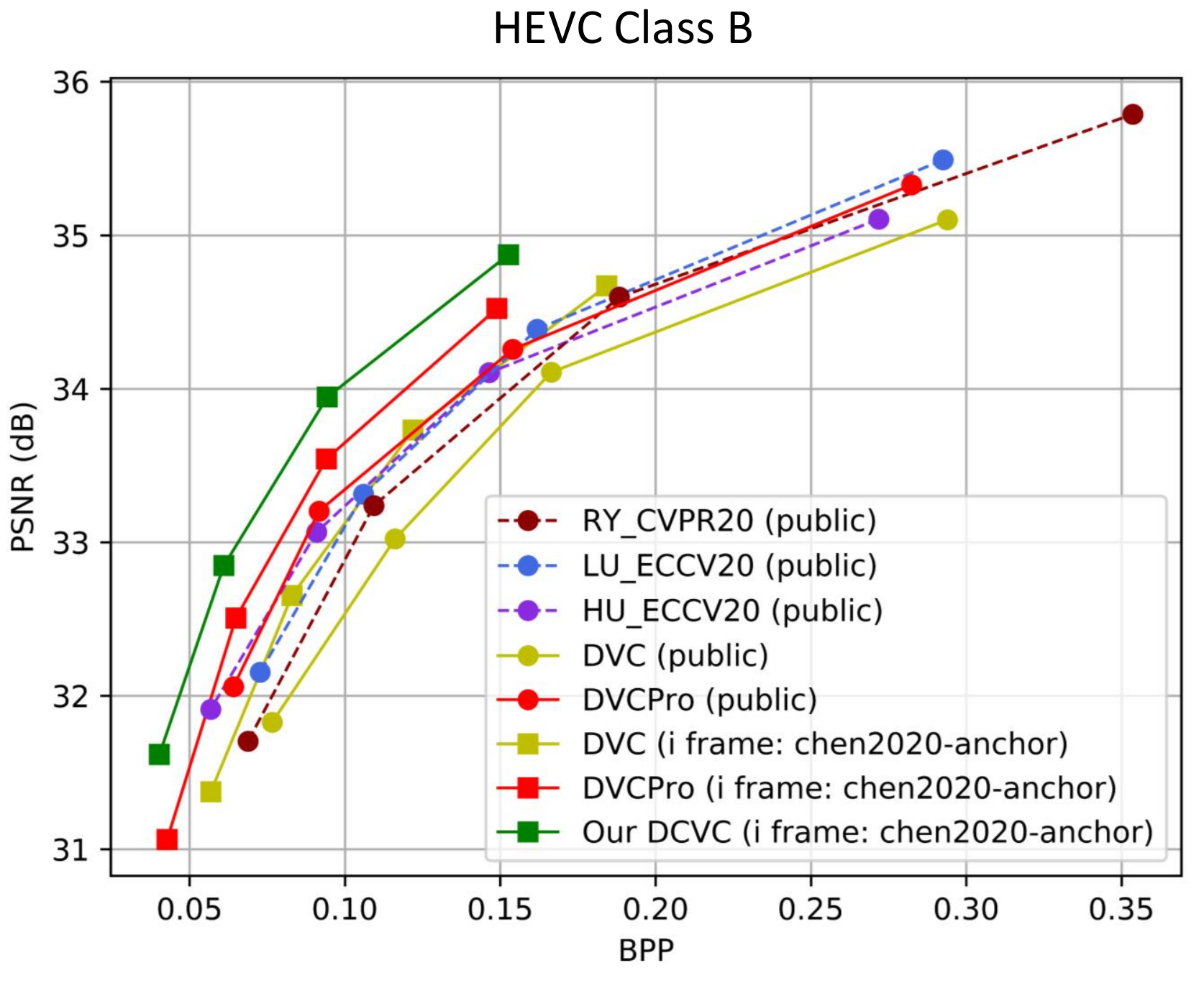}\\

     \caption{Performance comparison with the public results. The results of method with suffix *(public) are provided by \cite{TutorialVCIP, PyTorchVideoCompression}. The results of method with suffix *(i frame:  cheng2020-anchor) use  \textit{cheng2020-anchor} as the intra frame coding.
     }\label{rd_curve_public_comp}
 \end{figure}
    \subsection{Influence of  intra frame coding and comparison with more baselines}
        To build the best DL-based video compression framework, we use the SOTA DL-based image compression as our intra frame coding.   It is noted that, for fair comparison, DVC and DVCPro are retested using the same intra frame coding with DCVC, and their results are better than that reported in  their original papers.  Fig. \ref{rd_curve_public_comp} shows the performance comparison between our retested DVC/DVCPro and their  public results provided by \cite{TutorialVCIP, PyTorchVideoCompression}. In addition, Fig. \ref{rd_curve_public_comp} also shows the  results of the recent works RY\_CVPR20 \cite{Yang_2020_CVPR}, LU\_ECCV20 \cite{lu2020content}, and HU\_ECCV20\cite{hu2020improving}, provided  by \cite{TutorialVCIP, PyTorchVideoCompression}. From the public results in this figure, we can find that DVCPro is one SOTA  method among recent works. The results of LU\_ECCV20 and  HU\_ECCV20 are quite close with DVCPro.
    When the intra frame coding of DVC and DVCPro uses SOTA DL-based image compression model \textit{ cheng2020-anchor} provided by CompressAI \cite{begaint2020compressai}, their performance has large improvement. When using the same intra frame coding, our proposed DCVC method can significantly outperform DVCPro, as shown in Fig. \ref{rd_curve_public_comp}.

    \section{Discussion}
    \label{Sec_Discussion}
    In this paper, we make efforts on designing a conditional coding-based deep video compression framework which has a lower entropy bound than the commonly-used residue coding-based framework.  Residue coding-based framework assumes the inter frame prediction is always  most efficient, which is  inadequate, especially for encoding new contents.  By contrast, our conditional coding enables the  adaptability between learning  temporal correlation and learning spatial correlation.
    In addition, the condition is defined as feature domain context in DCVC. Context with higher dimensions can provide richer information to help the conditional coding, especially for the high frequency contents. In the future, high resolution video is more popular. High resolution video contains more high frequency contents, which means the advantage of our DCVC will be  more obvious.  

    When designing a conditional coding-based framework, the core questions are \textit{What is  condition? How to use  condition? And how to learn  condition?} In this paper, DCVC is a solution which answers these questions and demonstrates its effectiveness. However, these core questions are still open.
    Our DCVC framework is extensible and worthy more investigation. There exists great  potential in designing a more efficient solution by better defining, using and learning   the condition.

    In this paper, we do not add supervision on the channels in context during the training. There may  exist redundancy across channels, and this is not conducive to making full advantage of context with high dimensions. In the future, we will investigate how to  eliminate the redundancy across the channels  to maximize the utilization of context. For context generation, this paper only uses  single reference frame. Traditional  codecs have shown that using more reference frames can significantly improve the performance. Thus, how the design the conditional coding-based framework given multiple reference frames is very promising.  In addition, we currently do not consider temporal stability of reconstruction quality, which can be further improved by post processing or additional training supervision (e.g., loss about temporal stability).

\clearpage
\begin{appendices}
Appendices include detailed network structures, training strategies, as well as additional experimental results to demonstrate the effectiveness of the proposed DCVC.
\section{Network Architecture}

\begin{figure}[h]
    \centering
    \includegraphics[width=380pt, height=140pt]{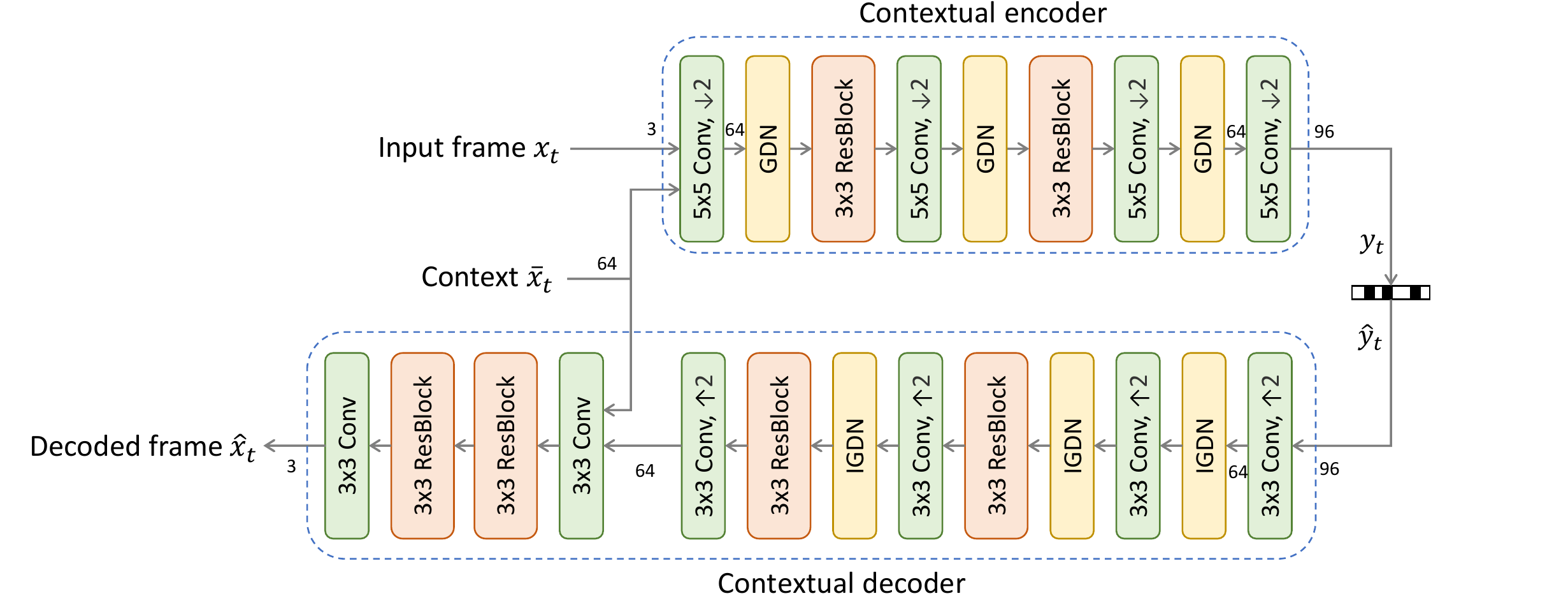}\\

    \caption{Network structure of contextual encoder and decoder. The above part is the encoder and the below part is the decoder. For simplification, the entropy model is omitted.  GDN is generalized divisive normalization \cite{ball2017endtoend} and IGDN is the inverse GDN. ResBlock represents plain residual block. The numbers represent channel dimensions.
    }\label{net_structure_CC}
\end{figure}

\begin{figure}[h]
    \centering
    \includegraphics[width=280pt, height=80pt]{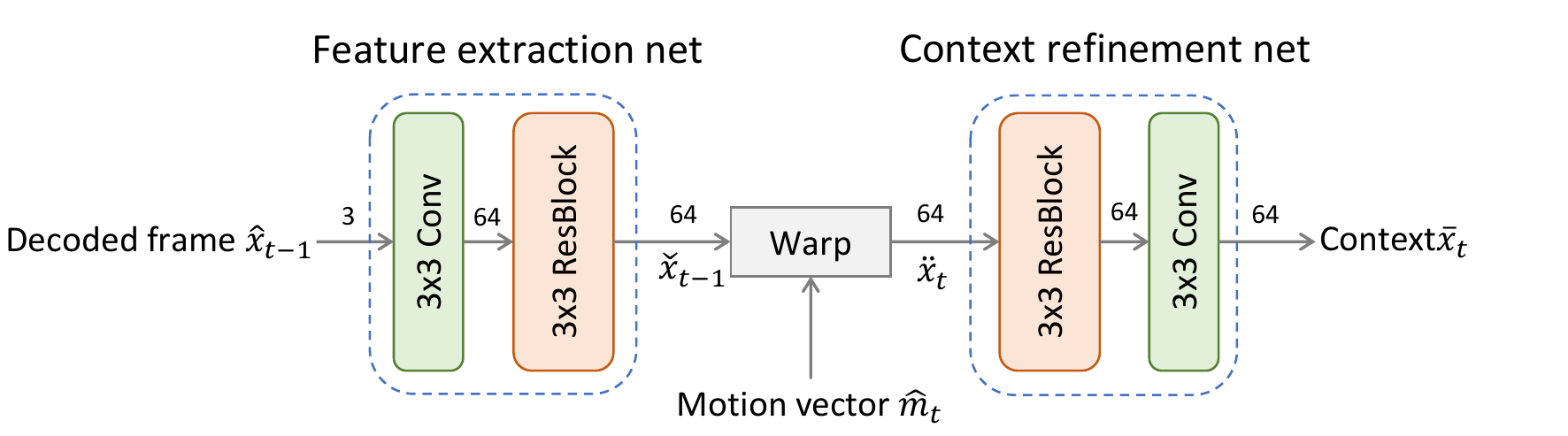}\\

    \caption{Network structure of   feature extraction network and context refinement network.
    }\label{net_structure_ref}
\end{figure}

\begin{figure}[h]
    \centering
    \includegraphics[width=280pt, height=65pt]{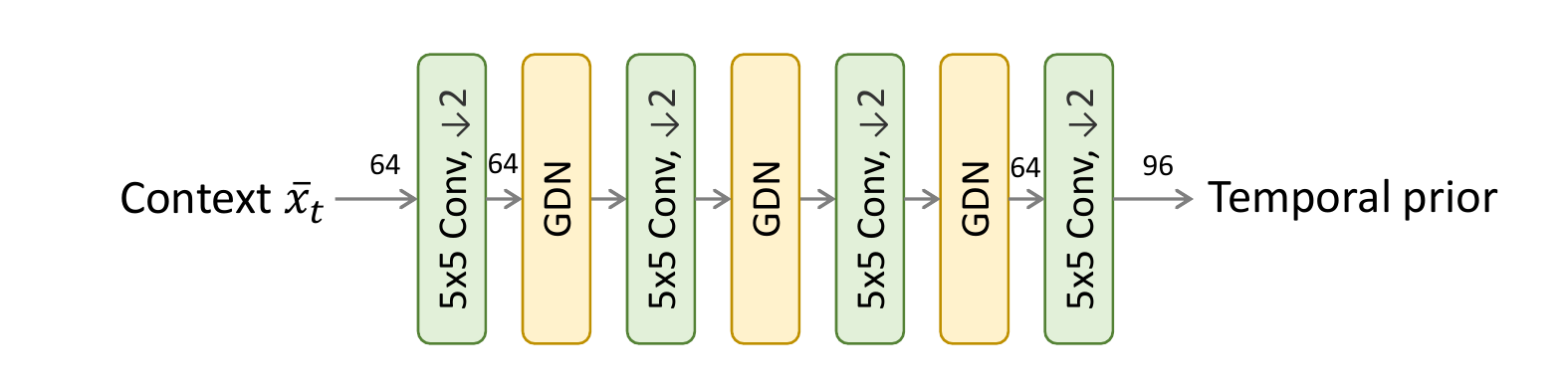}\\

    \caption{Network structure of   temporal prior encoder network, same with the commonly used encoder in image compression \cite{minnen2018joint} except the channel dimensions.
    }\label{net_structure_temporal_prior}
\end{figure}
\textbf{Contextual encoder and decoder}  The structures of our contextual encoder and decoder are illustrated in Fig. \ref{net_structure_CC}. For the contextual encoder, the input is the concatenation of the current frame $x_{t}$ and context $\bar{x}_{t}$. The  contextual encoder    encodes the concatenated data    into 16x down-sampled latent codes with dimension 96.
For the contextual decoder, we first up-sample the latent codes into the feature with original resolution. Then the up-sampled  feature  concatenated with context $\bar{x}_{t}$ is used to generate the final reconstruction frame $\hat{x}_{t}$.

\textbf{Feature extraction and context refinement} Fig. \ref{net_structure_ref} shows the structures of our feature  extraction network and context refinement network. Both of them   contains a convolution layer and a residual block. Consider the complexity, we do not use deeper network at present.

\textbf{Motion vector generation}  The motion vector (MV) generation part contains the motion estimation, MV encoder, and decoder.
For motion estimation, we use optical flow estimation network \cite{ranjan2017optical} to generate MV, like DVCPro \cite{lu2020end}. The network structures of MV encoder and decoder (decoder also contains a MV refine network) are same with those in DVCPro \cite{lu2020end}.

\textbf{Entropy model}  In the entropy model  for compressing the quantized latent codes $\hat{y}_{t}$, the  temporal prior encoder   network is borrowed from the encoder in image compression \cite{minnen2018joint} and consists of plain convolution layers (stride is set as 2 for down-sampling) and  GDN \cite{ball2017endtoend}, as shown in Fig. \ref{net_structure_temporal_prior}.  The hyper prior encoder/decoder, auto regressive network, and prior fusion network follow the entropy model in image compression \cite{minnen2018joint}. In addition, the  MV latent codes also have corresponding entropy model, where we  only use auto regressive model and hyper prior model. There is no temporal prior encoder in the the entropy model  for MV latent codes.

\section{Progressive training}

\begin{table}[]
    \centering
    \caption{The  training loss used in progressive training}
    \label{loss}
    {\def\arraystretch{1.2}
        \begin{adjustbox}{width=0.75\textwidth}
            \begin{tabular}{lll}
                \toprule
                Step & Loss & Calculation  \\
                \hline
                1&   $ L_{me}$&$ \lambda \cdot D(x_{t},\tilde{x}_{t})  +  R(\hat{g}_{t}) +  R(\hat{s}_{t})$  \\
                2&   $ L_{reconstruction}$&$\lambda \cdot D(x_{t},\hat{x}_{t})$  \\
                3&    $ L_{contextual\_coding}$&  $\lambda \cdot D(x_{t},\hat{x}_{t}) + R(\hat{y}_{t}) +  R(\hat{z}_{t})$\\
                4&    $ L_{all}$& $\lambda \cdot D(x_{t},\hat{x}_{t}) + R(\hat{y}_{t}) +  R(\hat{z}_{t}) +  R(\hat{g}_{t}) +  R(\hat{s}_{t})$ \\ \bottomrule
            \end{tabular}
        \end{adjustbox}
    }
\end{table}
The training loss consists of two metrics, i.e. the distortion $D$ and the bitrate cost $R$. In our method, the bitstream contains four parts, namely  $\hat{y}_{t}$, $\hat{g}_{t}$, $\hat{z}_{t}$, and $\hat{s}_{t}$.
$\hat{y}_{t}$ and $\hat{g}_{t}$ are the quantized latent codes of the current frame and MV, respectively.  $\hat{z}_{t}$ and $\hat{s}_{t}$ are their corresponding hyper priors. Thus, the total rate-distortion loss $ L_{all}$ should contain the bitrate costs of these four parts. The calculation manner of  $ L_{all}$ is shown in Table \ref{loss}.

While we already use the pre-trained optical flow estimation network \cite{ranjan2017optical} as the initialization of motion estimation,
the training may be still unstable if we directly use $L_{all}$ at the initial stage. Sometimes, the MV bitrate cost is very small but the total rate-distortion loss is large. This is because that the model thinks directly learning to generate context without MEMC  is  easier. Inspired by the existing work \cite{lin2020m} where the progressive training strategy is used,  we customize a progressive training strategy for our framework. The training is divided into four steps and the training loss for each step is shown in Table \ref{loss}:

\textbf{Step 1.} Warm up the  MV generation part  including motion estimation, MV encoder and decoder. The training loss is $ L_{me}$.  In Table \ref{loss}, $\tilde{x}_{t}$ is the warped frame in pixel domain, namely using $\hat{m}_{t}$ to do warping operation  on $\hat{x}_{t-1}$.

\textbf{Step 2.} Train other modules except the  MV generation part.  At this step, the parameters of MV generation part are frozen.  The training loss is $ L_{reconstruction}$. It means that we only pursue high reconstruction quality. This step is helpful for model to generate context which can better reconstruct the high frequency contents.

\textbf{Step 3.}   Based on previous step, the bit cost is considered, and the training loss becomes  $ L_{contextual\_coding}$. This step can be regarded as whole framework training with only freezing the MV generation part.

\textbf{Step 4.}  Reopen the MV generation part and perform the end-to-end training of whole framework according to   $ L_{all}$.

The proposed progressive training strategy can stabilize the model training.  For the training time, currently we need about one week on single Tesla V100 GPU. We will develop more advanced training technology to shorten the training time in the future. In addition, it is noted that we also apply the progressive strategy for  DVCPro (it does not have released models and we retrain it). For DVC, we just use the released models \cite{PyTorchVideoCompression}.

\section{Details of experimental settings}

\textbf{Dataset} The training dataset comes from Vimeo-90k septuplet  dataset \cite{xue2019video} (MIT License\footnote{https://github.com/anchen1011/toflow/blob/master/LICENSE}). The testing data includes MCL-JCV dataset \cite{wang2016mcl} (copyright can be found from this link \footnote{http://mcl.usc.edu/mcl-jcv-dataset/}), UVG dataset\cite{uvg} (BY-NC license\footnote{https://creativecommons.org/licenses/by-nc/3.0/deed.en\_US}), and HEVC standard test videos (more details can be found in \cite{bossen2013common}). These datasets are commonly-used for  video compression research and can be downloaded from Internet. The consents of these datasets are public. In addition, we have manually checked that these datasets do not contain personally identifiable information or offensive content.

\textbf{Intra frame coding}
The intra frame coding in our framework  directly uses the existing deep image compression models,  where the model parameters are provided by CompressAI \cite{begaint2020compressai} (Apache License 2.0 \footnote{https://github.com/InterDigitalInc/CompressAI/blob/master/LICENSE}).
We use \textit{ cheng2020-anchor} \cite{cheng2020learned} for MSE target and use \textit{hyperprior} \cite{balle2018variational} for MS-SSIM target, as they are the best models provided by CompressAI.

In DCVC, we train 4 models with different  $\lambda$ s \{MSE: 256, 512, 1024, 2048; MS-SSIM: 8, 16, 32, 64\}.  The models with   quality index 3, 4, 5, 6 (trained with 4 different $\lambda$ s)  in CompressAI are used for the corresponding intra frame coding. For example,  the model with   quality index 6 in CompressAI is used for our DCVC model with $\lambda$ 2048 (for MSE target) or 64 (for MS-SSIM target).

\textbf{FFMPEG settings} We test the x264 and x265 encoders from FFMPEG\cite{FFMPEG}.
The settings of  these two encoders are same with \cite{lu2020end} except two options. One is  that we use the \textit{veryslow} preset  rather than \textit{veryfast} preset.   \textit{Veryslow} preset can achieve higher compression ratio   than \textit{veryfast} preset. Another is that we use the constant quantization parameter setting rather than constant  rate factor  setting, where constant quantization parameter setting can avoid the influence of rate control.
The detailed configurations  of x264 and x265 are
\begin{itemize}
    \item  x264: \textit{ffmpeg -pix fmt yuv420p -s WxH -r FR -i Video.yuv -vframes N -c:v libx264 -preset veryslow  -tune zerolatency -qp QP -g GOP -bf 2 -b strategy 0 -sc threshold 0 output.mkv}
    \item x265:   \textit{ffmpeg -pix fmt yuv420p -s WxH -r FR -i Video.yuv -vframes N -c:v libx265 -preset veryslow  -tune zerolatency -x265-params ”qp=QP:keyint=GOP” output.mkv}
\end{itemize}
\textit{W, H, FR, N, QP} and \textit{GOP} represent the width, height, frame rate, the number of encoded frames,  quantization parameter, and group of pictures, respectively.

\textbf{GOP size and tested frame number} We follow \cite{lu2020end} and set the GOP size as   10 for HEVC test videos and 12 for non-HEVC test  videos, respectively. The  tested frame number of HEVC videos is 100 (10 GOPs), same with  \cite{lu2020end}. As there is no description about the tested frame number for non-HEVC test videos in \cite{lu2020end}, we test 120 frames for MCL-JCV and UVG datasets, which has 10 GOPs, same with  HEVC test videos.


\section{Test on larger GOP size}
\begin{figure}[h]
    \centering
    \includegraphics[width=250pt, height=150pt]{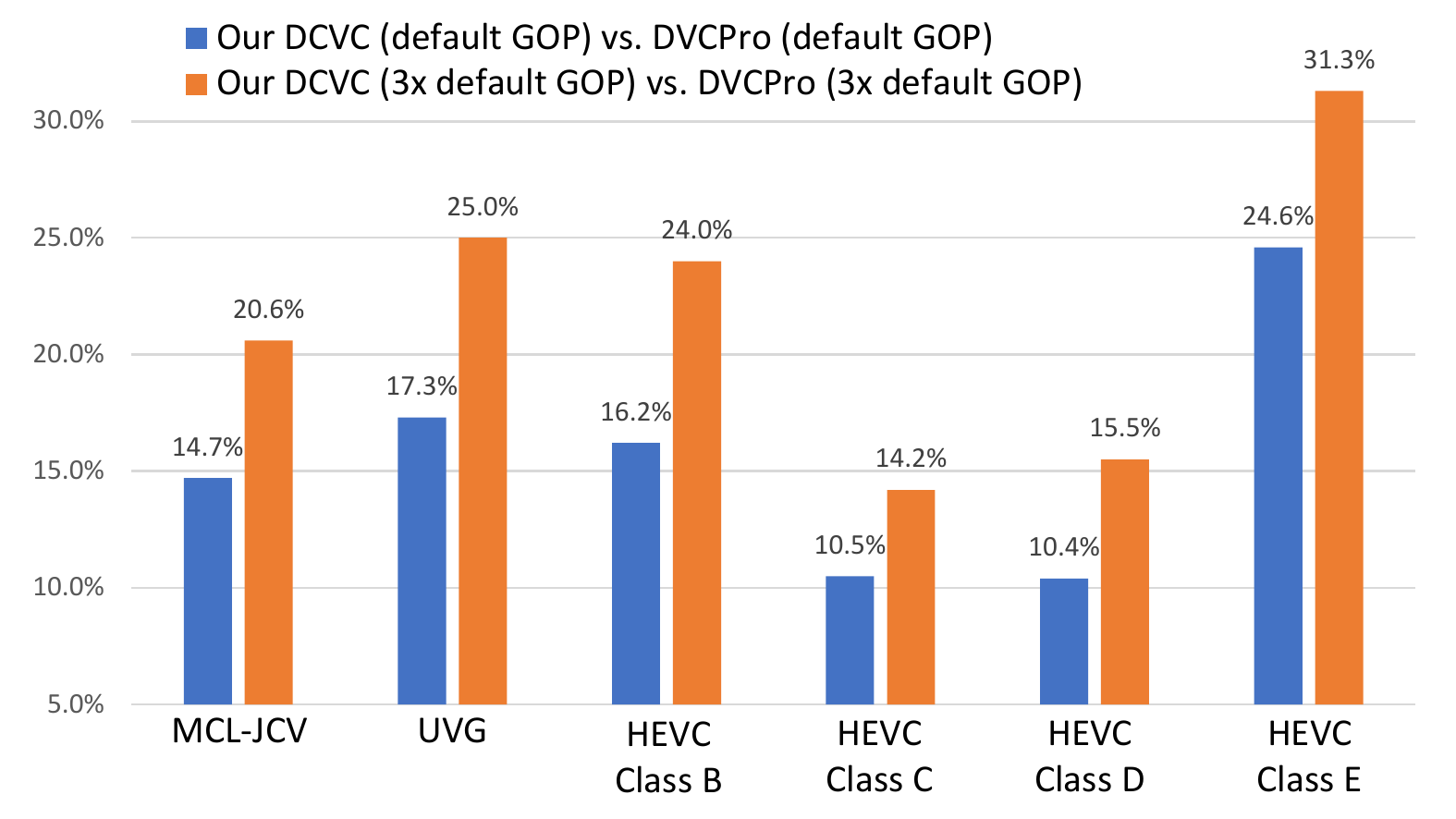}\\

    \caption{Bitrate saving under different GOP settings. Default GOP setting is \{HEVC test videos: 10, non-HEVC test videos: 12\}, same with \cite{lu2020end}. 3x default GOP setting is \{HEVC test videos: 30, non-HEVC test videos: 36\}. The tested frame number under two GOP settings is \{HEVC test videos: 30, non-HEVC test videos: 36\}. We can find that the bitrate saving of our DCVC is larger under larger GOP size.
    }\label{rd_curve_diffGOP}
\end{figure}
\begin{figure}[h]
    \centering
    \includegraphics[width=280pt, height=170pt]{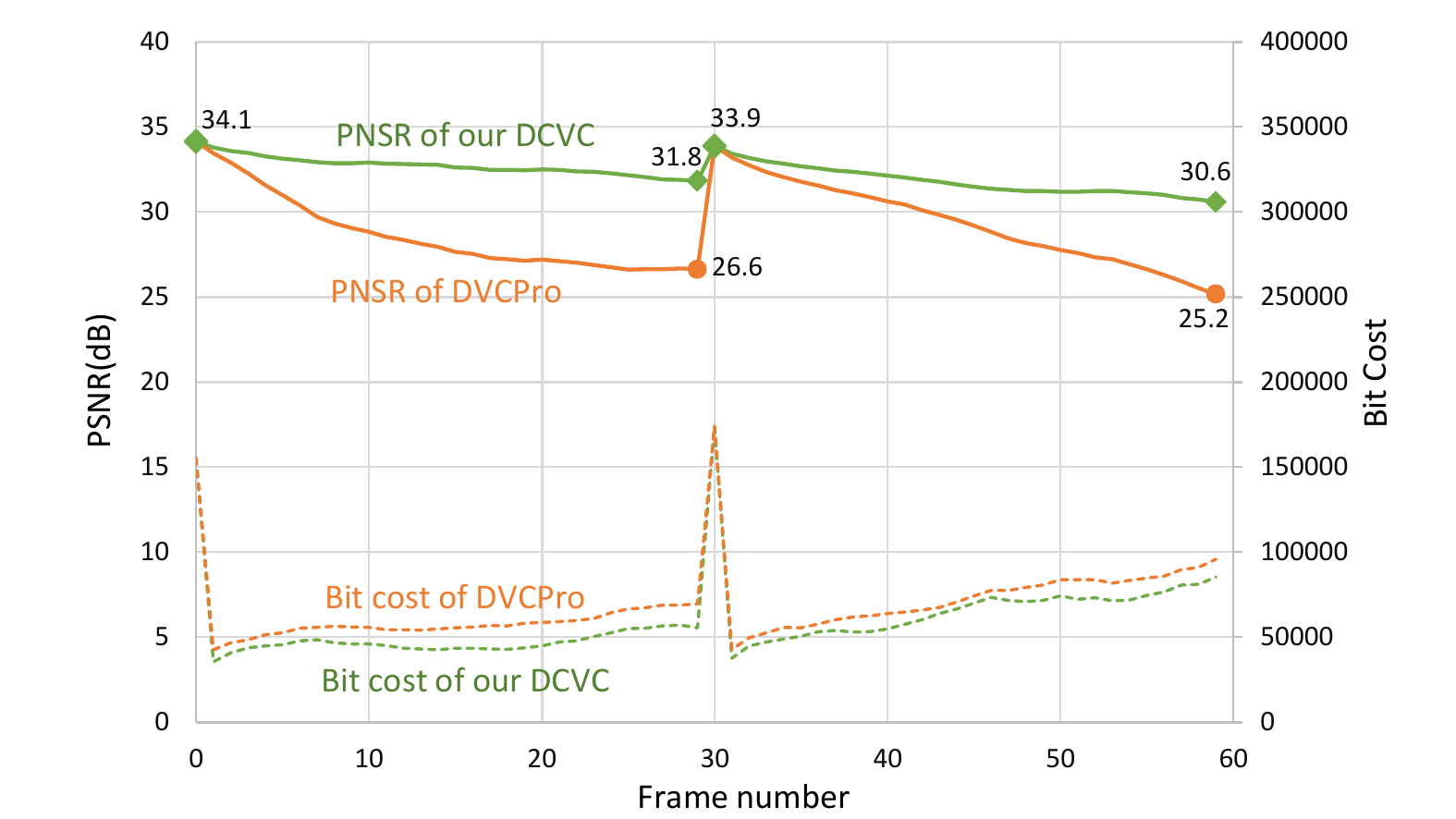}\\

    \caption{Example of PSNR and bit cost comparison between our DCVC and DVCPro.  The tested video is \textit{BasketballDrive} from HEVC Class B dataset. The GOP size is 30, and tested frame number is 60. From this example, we can find that our DCVC can efficiently alleviate the error-propagation problem. DCVC can use fewer bits while achieving much better reconstruction quality.
    }\label{psnr_curve_example}
\end{figure}
In the paper, we  follow  \cite{lu2020end} and set the GOP size as   10 for HEVC test videos and 12 for non-HEVC test video, denoted as default GOP setting.  Actually, this GOP setting is relatively small when compared that in practical scenarios. For this reason, we conduct the experiments under larger GOP size. The bitrate saving comparison is shown in Fig. \ref{rd_curve_diffGOP}. In this comparison, we increase the GOP size to 3 times of default GOP setting,  i.e. 30   for HEVC test videos and 36 for non-HEVC test videos. From this comparison, we can find that, when compared with DVCPro, the improvement of our DCVC is much larger under 3x default GOP size.
It shows that our conditional coding-based framework can better deal with the error-propagation problem. Under large GOP size, residue coding still  assumes that the   inter frame prediction is always  most efficient even when the quality of reference frame is bad, then  suffers from the large prediction error. By contrast, our conditional coding does not need to pursue the strict equality between prediction frame and the current frame, and enables the  adaptability between learning  temporal correlation and learning spatial correlation. Thus, the advantage of our DCVC will be  more obvious when the GOP size increases.
In addition, the bitrate saving increase is larger for high resolution videos. For example, for the 240P dataset HEVC Class D, the bitrate saving is changed from 10.4\% to 15.5\%. By contrast,  for the 1080P dataset HEVC Class B,   the bitrate saving is changed from 16.2\% to 24.0\%. It is because that the context in  feature domain  is helpful for reconstructing the high frequency contents, then the reconstruction quality can be improved and it is  conducive  to alleviating error-propagation problem.

An example of PSNR and bit cost comparison between our DCVC and DVCPro is shown in Fig. \ref{psnr_curve_example}. In the example, the PSNR of DVCPro decreases from 34.1 dB to 26.6 dB in the first GOP. By contrast, our DCVC only decreases to 31.8 dB.

\section{Ablation study}

\begin{figure}
    \centering
    \includegraphics[width=300pt, height=170pt]{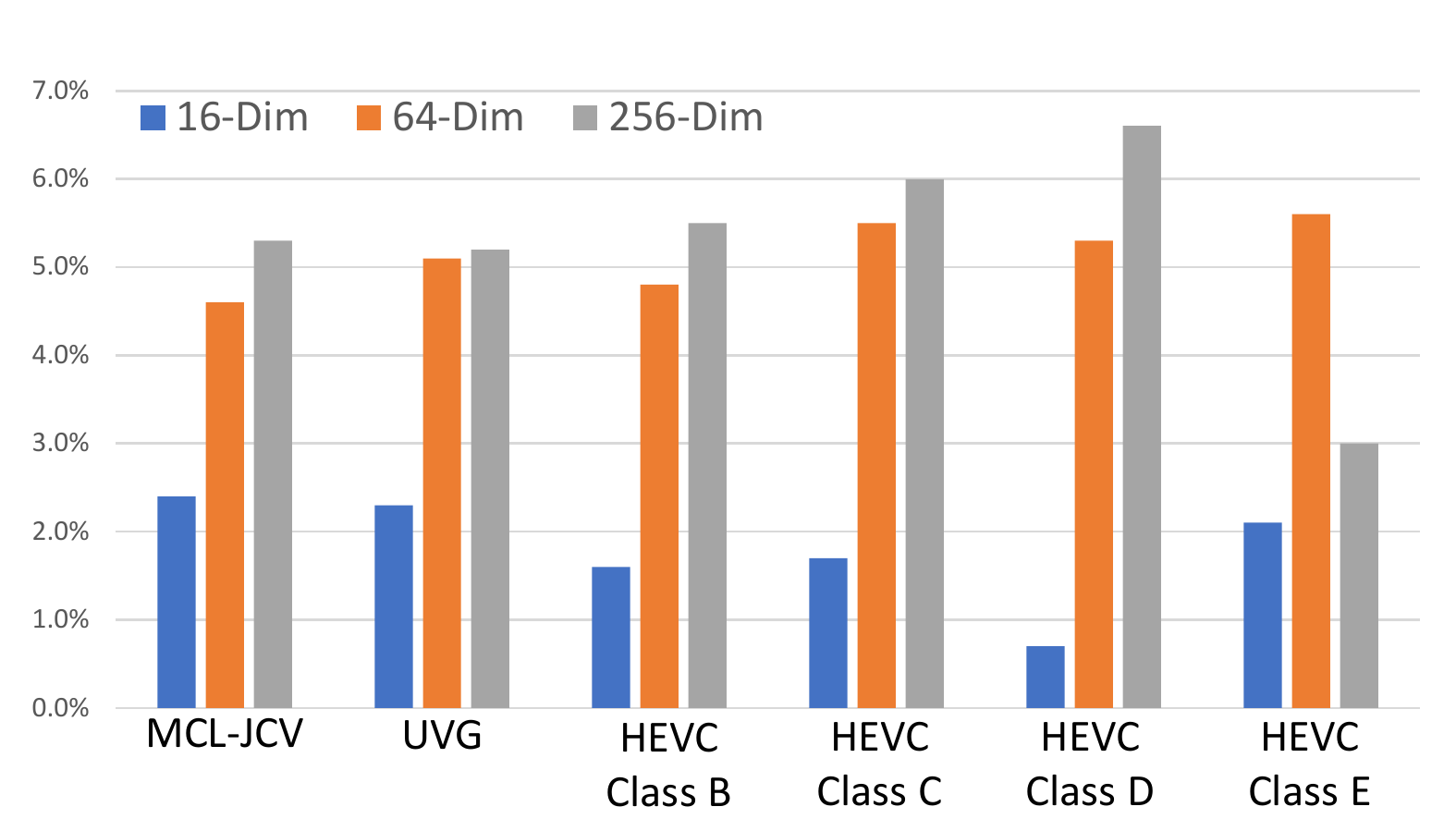}
    \captionof{figure}{Bitrate saving when using different channel dimensions for context. The anchor is 3-Dim (dimension is 3) model.}
    \label{result_channelNum}
\end{figure}
\textbf{Channel dimension of context} In  DCVC, the channel dimension of context is set as 64 in the implementation. We also conduct the experiment when different dimensions are used. The cases (3, 16, 256-Dim) are also tested. The corresponding bitrate saving comparison is shown in Fig. \ref{result_channelNum}, where the anchor is the  3-Dim model. From this figure, we can observe that the 16-Dim model can improve the performance in some degree, and the 64-Dim model can further boost the performance in a larger degree for most datasets. However, the improvement brought by  256-Dim model is relatively small. The HEVC Class E even has performance loss. The reason may be that the model training is not stable if there is no extra supervision for training the context with so high dimensions and in original resolution. For this reason, we  adopt the 64-Dim model  at present.

\begin{figure}
    \centering
    \includegraphics[width=240pt, height=190pt]{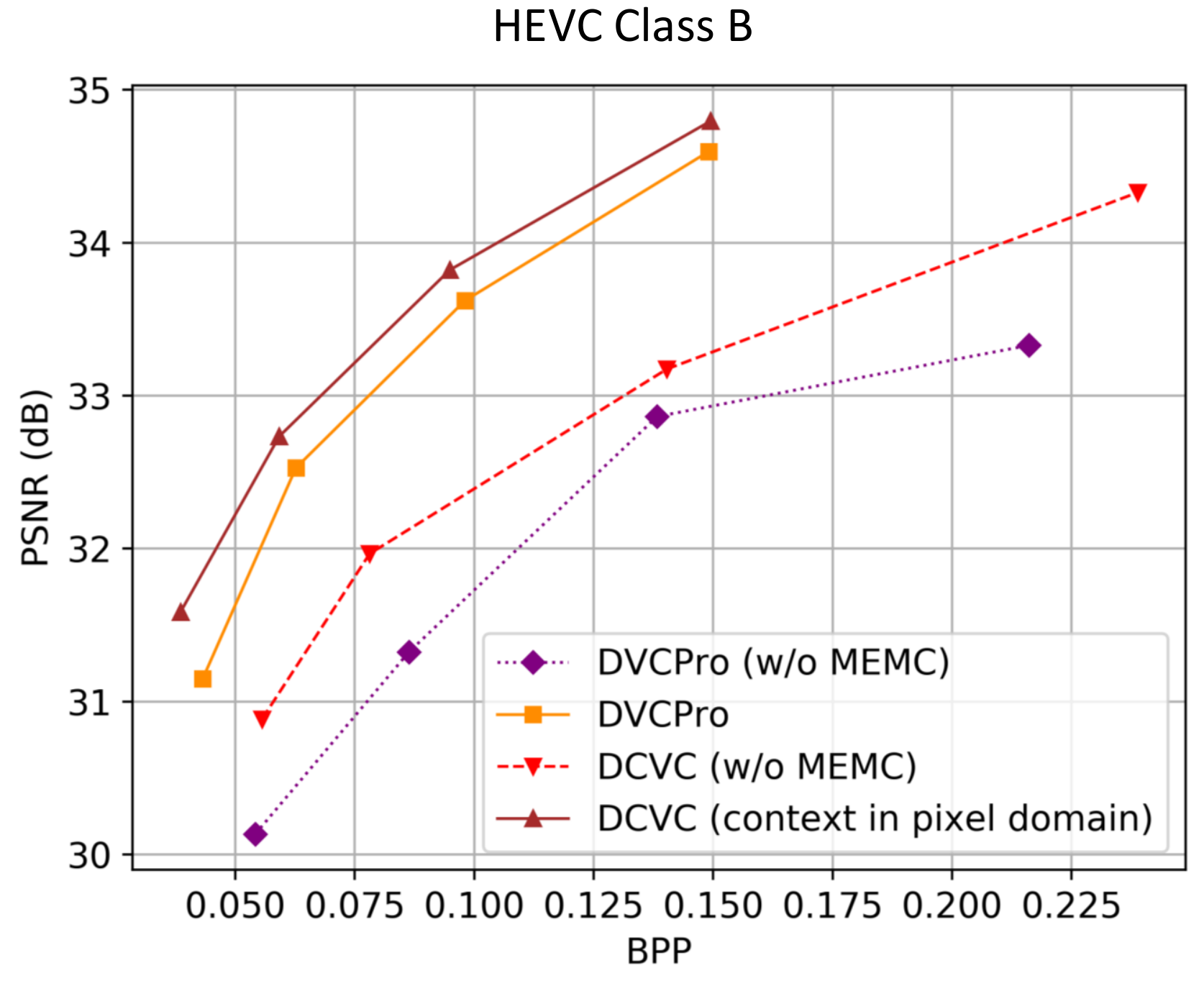}\\

    \caption{ Performance comparison when MEMC is disabled. DCVC (context in pixel domain) refers the model using  temporal prior and concatenating RGB prediction.
    }\label{rd_curve_noMV}
\end{figure}
\textbf{Motion estimation and motion compensation (MEMC)} In our DCVC, we  use MEMC to guide the model where to extract context. Actually we are also very interested in the case without MEMC. For these reason, we test the DCVC and DVCPro where the MEMC is removed (directly use the previous decoded frame as the predicted frame in DVCPro and the condition in DCVC). They are denoted as DVCPro (w/o MEMC) and DCVC (w/o MEMC), respectively.  As DCVC (w/o MEMC) uses the previous decoded frame  as condition, we use the model DCVC (context in pixel domain, i.e.  temporal prior + concatenating RGB prediction) for fair comparison.
Fig. \ref{rd_curve_noMV} shows the results. From this figure, we can find that the performance has a large drop for both of    DVCPro and DCVC  if MEMC is removed. However,  DCVC (w/o MEMC) is still better than DVCPro (w/o MEMC), and the performance gap is even larger. The DCVC (context in pixel domain) has 12.7\% improvement over DVCPro. By contrast, DCVC (w/o MEMC) can achieve 22.1\% bitrate saving compared with  DVCPro (w/o MEMC). These results show that the MEMC is helpful for both  frame residue coding and conditional coding-based frameworks.  When  MEMC is disable, the improvement of conditional coding can be larger.

While we currently  use MEMC to learn the context, there still exists great potential in designing a better learning manner. In the future, we will continue  the investigation. For example,  transformer \cite{khan2021transformers} can be used to explore the global correlations and generate the context with larger receptive field.

\section{Visual comparison}

\begin{figure}
    \centering
    \includegraphics[width=400pt, height=320pt]{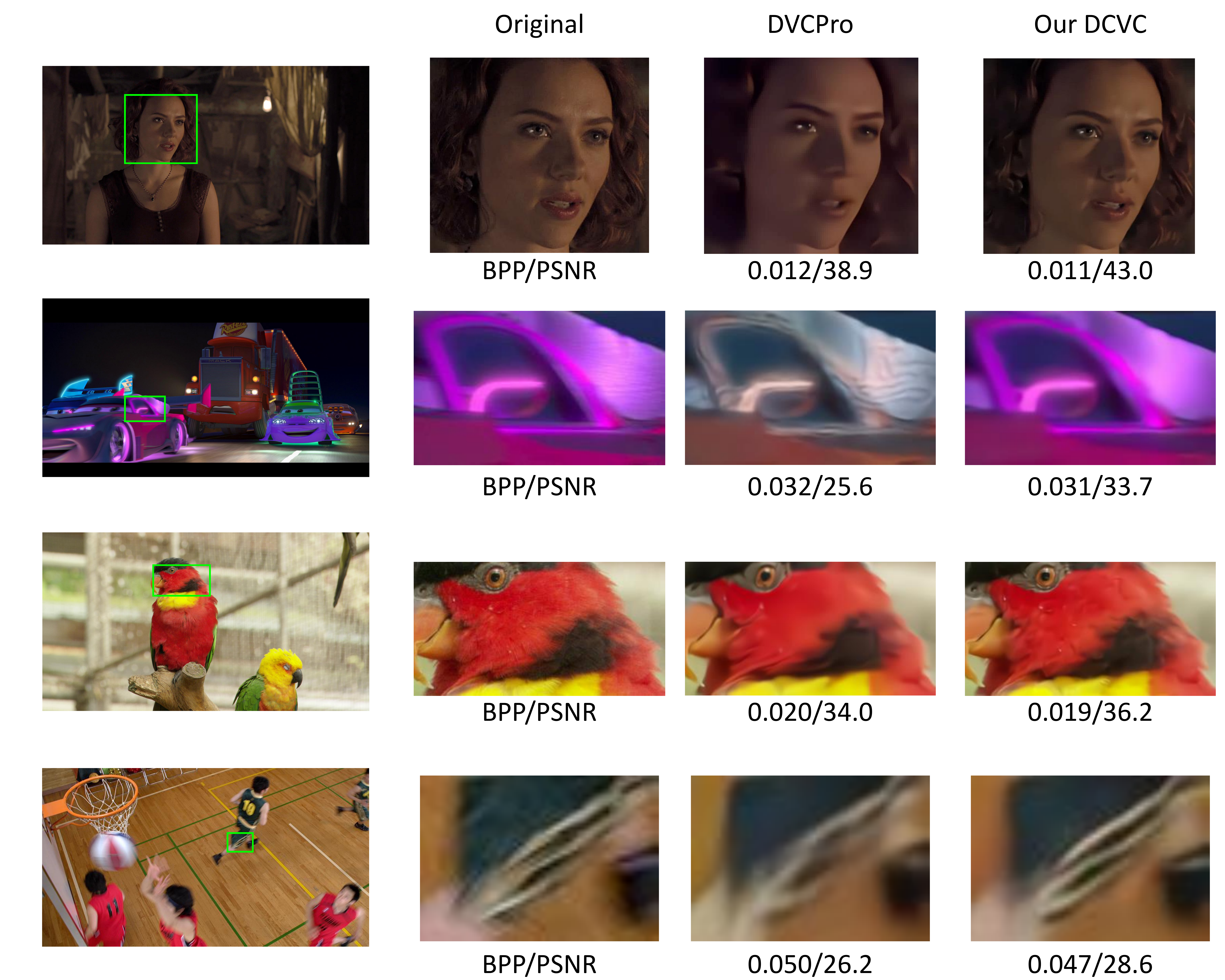}\\

    \caption{Examples of visual comparison. The first column shows the original full frames. The second column shows the cropped patch in original frame. The contents in third and fourth columns are reconstructed by DVCPro and our DCVC, respectively.
    }\label{visual_example}
\end{figure}
We also conduct the visual comparison between the previous SOTA DVCPro and our DCVC. Several examples are shown in Fig. \ref{visual_example}.
From these examples, we can find that our DCVC can achieve much higher reconstruction quality without increasing the bitrate cost. For instance, in the example shown in the second row in  Fig. \ref{visual_example}, we can find that the image reconstructed by DVCPro has obvious color  distortion and unexpected textures. By contrast, our DCVC can achieve much better  results.  In the example shown in the fourth row, our DCVC also  produces much clearer stripe texture in the basketball clothes.

\end{appendices}


\begin{thebibliography}{10}

\bibitem{girod1995comparison}
B.~Girod, E.~G. Steinbach, and N.~Faerber, ``{Comparison of the H. 263 and H.
  261 video compression standards},'' in {\em Standards and Common Interfaces
  for Video Information Systems: A Critical Review}, 1995.

\bibitem{bross2021developments}
B.~Bross, J.~Chen, J.-R. Ohm, G.~J. Sullivan, and Y.-K. Wang, ``{Developments
  in international video coding standardization after AVC, with an overview of
  Versatile Video Coding (VVC)},'' {\em Proceedings of the IEEE}, 2021.

\bibitem{lu2019dvc}
G.~Lu, W.~Ouyang, D.~Xu, X.~Zhang, C.~Cai, and Z.~Gao, ``{DVC: an end-to-end
  deep video compression framework},'' in {\em Proceedings of the IEEE/CVF
  Conference on Computer Vision and Pattern Recognition}, pp.~11006--11015,
  2019.

\bibitem{lu2020end}
G.~Lu, X.~Zhang, W.~Ouyang, L.~Chen, Z.~Gao, and D.~Xu, ``An end-to-end
  learning framework for video compression,'' {\em IEEE transactions on pattern
  analysis and machine intelligence}, 2020.

\bibitem{lu2020content}
G.~Lu, C.~Cai, X.~Zhang, L.~Chen, W.~Ouyang, D.~Xu, and Z.~Gao, ``Content
  adaptive and error propagation aware deep video compression,'' in {\em
  European Conference on Computer Vision}, pp.~456--472, Springer, 2020.

\bibitem{Yang_2020_CVPR}
R.~Yang, F.~Mentzer, L.~V. Gool, and R.~Timofte, ``Learning for video
  compression with hierarchical quality and recurrent enhancement,'' in {\em
  Proceedings of the IEEE/CVF Conference on Computer Vision and Pattern
  Recognition}, 2020.

\bibitem{lin2020m}
J.~Lin, D.~Liu, H.~Li, and F.~Wu, ``{M-LVC}: multiple frames prediction for
  learned video compression,'' in {\em Proceedings of the IEEE/CVF Conference
  on Computer Vision and Pattern Recognition}, 2020.

\bibitem{Djelouah_2019_ICCV}
A.~Djelouah, J.~Campos, S.~Schaub-Meyer, and C.~Schroers, ``Neural inter-frame
  compression for video coding,'' in {\em Proceedings of the IEEE/CVF
  International Conference on Computer Vision (ICCV)}, October 2019.

\bibitem{agustsson2020scale}
E.~Agustsson, D.~Minnen, N.~Johnston, J.~Balle, S.~J. Hwang, and G.~Toderici,
  ``Scale-space flow for end-to-end optimized video compression,'' in {\em
  Proceedings of the IEEE/CVF Conference on Computer Vision and Pattern
  Recognition}, pp.~8503--8512, 2020.

\bibitem{wu2018vcii}
C.-Y. Wu, N.~Singhal, and P.~Kr{\"a}henb{\"u}hl, ``Video compression through
  image interpolation,'' in {\em ECCV}, 2018.

\bibitem{yang2020hierarchical}
R.~Yang, Y.~Yang, J.~Marino, and S.~Mandt, ``Hierarchical autoregressive
  modeling for neural video compression,'' {\em 9th International Conference on
  Learning Representations, {ICLR}}, 2021.

\bibitem{ladune2020optical}
T.~Ladune, P.~Philippe, W.~Hamidouche, L.~Zhang, and O.~D{\'{e}}forges,
  ``Optical flow and mode selection for learning-based video coding,'' in {\em
  22nd {IEEE} International Workshop on Multimedia Signal Processing}, 2020.

\bibitem{balle2018variational}
J.~Ball{\'e}, D.~Minnen, S.~Singh, S.~J. Hwang, and N.~Johnston, ``Variational
  image compression with a scale hyperprior,'' {\em 6th International
  Conference on Learning Representations, {ICLR}}, 2018.

\bibitem{minnen2018joint}
D.~Minnen, J.~Ball{\'e}, and G.~Toderici, ``Joint autoregressive and
  hierarchical priors for learned image compression,'' {\em arXiv preprint
  arXiv:1809.02736}, 2018.

\bibitem{liu2020conditional}
J.~Liu, S.~Wang, W.-C. Ma, M.~Shah, R.~Hu, P.~Dhawan, and R.~Urtasun,
  ``Conditional entropy coding for efficient video compression,'' {\em arXiv
  preprint arXiv:2008.09180}, 2020.

\bibitem{rippel2019learned}
O.~Rippel, S.~Nair, C.~Lew, S.~Branson, A.~G. Anderson, and L.~Bourdev,
  ``Learned video compression,'' in {\em Proceedings of the IEEE/CVF
  International Conference on Computer Vision}, pp.~3454--3463, 2019.

\bibitem{ladune2021conditional}
T.~Ladune, P.~Philippe, W.~Hamidouche, L.~Zhang, and O.~D{\'e}forges,
  ``Conditional coding for flexible learned video compression,'' in {\em Neural
  Compression: From Information Theory to Applications -- Workshop @ ICLR},
  2021.

\bibitem{theis2017lossy}
L.~Theis, W.~Shi, A.~Cunningham, and F.~Husz{\'a}r, ``Lossy image compression
  with compressive autoencoders,'' {\em arXiv preprint arXiv:1703.00395}, 2017.

\bibitem{ball2017endtoend}
J.~Ballé, V.~Laparra, and E.~P. Simoncelli, ``End-to-end optimized image
  compression,'' {\em arXiv preprint arXiv:1611.01704}, 2017.

\bibitem{cheng2020learned}
Z.~Cheng, H.~Sun, M.~Takeuchi, and J.~Katto, ``Learned image compression with
  discretized gaussian mixture likelihoods and attention modules,'' in {\em
  Proceedings of the IEEE/CVF Conference on Computer Vision and Pattern
  Recognition}, pp.~7939--7948, 2020.

\bibitem{toderici2015variable}
G.~Toderici, S.~M. O'Malley, S.~J. Hwang, D.~Vincent, D.~Minnen, S.~Baluja,
  M.~Covell, and R.~Sukthankar, ``Variable rate image compression with
  recurrent neural networks,'' {\em arXiv preprint arXiv:1511.06085}, 2015.

\bibitem{toderici2017full}
G.~Toderici, D.~Vincent, N.~Johnston, S.~Jin~Hwang, D.~Minnen, J.~Shor, and
  M.~Covell, ``Full resolution image compression with recurrent neural
  networks,'' in {\em Proceedings of the IEEE Conference on Computer Vision and
  Pattern Recognition}, pp.~5306--5314, 2017.

\bibitem{johnston2018improved}
N.~Johnston, D.~Vincent, D.~Minnen, M.~Covell, S.~Singh, T.~Chinen, S.~J.
  Hwang, J.~Shor, and G.~Toderici, ``Improved lossy image compression with
  priming and spatially adaptive bit rates for recurrent networks,'' in {\em
  Proceedings of the IEEE Conference on Computer Vision and Pattern
  Recognition}, pp.~4385--4393, 2018.

\bibitem{pessoa2020end}
J.~Pessoa, H.~Aidos, P.~Tom{\'a}s, and M.~A. Figueiredo, ``End-to-end learning
  of video compression using spatio-temporal autoencoders,'' in {\em 2020 IEEE
  Workshop on Signal Processing Systems (SiPS)}, pp.~1--6, IEEE, 2020.

\bibitem{habibian2019video}
A.~Habibian, T.~v. Rozendaal, J.~M. Tomczak, and T.~S. Cohen, ``Video
  compression with rate-distortion autoencoders,'' in {\em Proceedings of the
  IEEE/CVF International Conference on Computer Vision}, pp.~7033--7042, 2019.

\bibitem{hu2020improving}
Z.~Hu, Z.~Chen, D.~Xu, G.~Lu, W.~Ouyang, and S.~Gu, ``Improving deep video
  compression by resolution-adaptive flow coding,'' in {\em European Conference
  on Computer Vision}, pp.~193--209, Springer, 2020.

\bibitem{bao2019depth}
W.~Bao, W.-S. Lai, C.~Ma, X.~Zhang, Z.~Gao, and M.-H. Yang, ``Depth-aware video
  frame interpolation,'' in {\em Proceedings of the IEEE/CVF Conference on
  Computer Vision and Pattern Recognition}, pp.~3703--3712, 2019.

\bibitem{niklaus2020softmax}
S.~Niklaus and F.~Liu, ``Softmax splatting for video frame interpolation,'' in
  {\em Proceedings of the IEEE/CVF Conference on Computer Vision and Pattern
  Recognition}, pp.~5437--5446, 2020.

\bibitem{wang2016mcl}
H.~Wang, W.~Gan, S.~Hu, J.~Y. Lin, L.~Jin, L.~Song, P.~Wang, I.~Katsavounidis,
  A.~Aaron, and C.-C.~J. Kuo, ``{MCL-JCV: a JND-based H. 264/AVC video quality
  assessment dataset},'' in {\em 2016 IEEE International Conference on Image
  Processing (ICIP)}, pp.~1509--1513, IEEE, 2016.

\bibitem{shannon2001mathematical}
C.~E. Shannon, ``A mathematical theory of communication,'' {\em ACM SIGMOBILE
  mobile computing and communications review}, vol.~5, no.~1, pp.~3--55, 2001.

\bibitem{PyTorchVideoCompression}
``{PyTorchVideoCompression}.''
  \url{https://github.com/ZhihaoHu/PyTorchVideoCompression}.
\newblock Online; accessed 12 April 2021.

\bibitem{ranjan2017optical}
A.~Ranjan and M.~J. Black, ``Optical flow estimation using a spatial pyramid
  network,'' in {\em Proceedings of the IEEE conference on computer vision and
  pattern recognition}, pp.~4161--4170, 2017.

\bibitem{xue2019video}
T.~Xue, B.~Chen, J.~Wu, D.~Wei, and W.~T. Freeman, ``Video enhancement with
  task-oriented flow,'' {\em International Journal of Computer Vision (IJCV)},
  vol.~127, no.~8, pp.~1106--1125, 2019.

\bibitem{bossen2013common}
F.~Bossen {\em et~al.}, ``Common test conditions and software reference
  configurations,'' in {\em JCTVC-L1100}, vol.~12, 2013.

\bibitem{uvg}
``Ultra video group test sequences.'' \url{http://ultravideo.cs.tut.fi}.
\newblock Online; accessed 12 April 2021.

\bibitem{begaint2020compressai}
J.~B{\'e}gaint, F.~Racap{\'e}, S.~Feltman, and A.~Pushparaja, ``{CompressAI: a
  PyTorch library and evaluation platform for end-to-end compression
  research},'' {\em arXiv preprint arXiv:2011.03029}, 2020.

\bibitem{vcip_XuLYT20}
D.~Xu, G.~Lu, R.~Yang, and R.~Timofte, ``Learned image and video compression
  with deep neural networks,'' in {\em 2020 {IEEE} International Conference on
  Visual Communications and Image Processing, {VCIP} 2020, Macau, China,
  December 1-4, 2020}, pp.~1--3, {IEEE}, 2020.

\bibitem{TutorialVCIP}
D.~Xu, G.~Lu, R.~Yang, and R.~Timofte, ``Tutorial: Learned image and video
  compression with deep neural networks.''
  \url{https://drive.google.com/file/d/162omgk0CmHPBj4J7vWsNr8N9SPn5j97F/view}.
\newblock Online; accessed 12 April 2021.

\bibitem{FFMPEG}
``Ffmpeg.'' \url{https://www.ffmpeg.org/}.
\newblock Online; accessed 12 April 2021.

\bibitem{bjontegaard2001calculation}
G.~Bjontegaard, ``{Calculation of average PSNR differences between
  RD-curves},'' {\em VCEG-M33}, 2001.

\bibitem{khan2021transformers}
S.~Khan, M.~Naseer, M.~Hayat, S.~W. Zamir, F.~S. Khan, and M.~Shah,
  ``Transformers in vision: A survey,'' {\em arXiv preprint arXiv:2101.01169},
  2021.

\end{thebibliography}
\end{document}